\documentstyle[epsf,psfig,emulatepasa]{article}

\def\reference{\parskip 0pt\par\noindent\hangindent 0.5 truecm}

\def\spose#1{\hbox to 0pt{#1\hss}}
\def\simlt{\mathrel{\spose{\lower 3pt\hbox{$\mathchar"218$}}
     \raise 2.0pt\hbox{$\mathchar"13C$}}}
\def\simgt{\mathrel{\spose{\lower 3pt\hbox{$\mathchar"218$}}
     \raise 2.0pt\hbox{$\mathchar"13E$}}}

\begin{document}

\small
\shorttitle{Galactic Chemical Evolution}
\shortauthor{B.K.~Gibson et~al.}
\title{\large \bf
Galactic Chemical Evolution
}
\author{\small 
 Brad K. Gibson, Yeshe Fenner, Agostino Renda, 
Daisuke Kawata \& Hyun-chul Lee
}

\date{}
\twocolumn[
\maketitle
\vspace{-20pt}
\small
{\center
Centre for Astrophysics \& Supercomputing, 
Swinburne University, Mail \#31, Hawthorn, VIC 3122, Australia\\
bgibson,yfenner,arenda,dkawata,hclee@astro.swin.edu.au\\[3mm]
}

\begin{center}
{\bfseries Abstract}
\end{center}
\begin{quotation}
\begin{small}
\vspace{-5pt}
The primary present-day observables upon which theories of galaxy
evolution are based are a system's morphology, dynamics, colour, and
chemistry. Individually, each provides an important constraint to any
given model; in concert, the four represent a fundamental (intractable)
boundary condition for chemodynamical simulations.  We review the current
state-of-the-art semi-analytical and chemodynamical models for the Milky
Way, emphasising the strengths and weaknesses of both approaches.
\end{small}
\end{quotation}
]

\bigskip

\section{Introduction}

Stars and interstellar gas in galaxies exhibit diverse chemical element
abundance patterns that are shaped by their environment and formation
histories. The aim of Galactic Chemical Evolution (GCE) is to use the observed
abundances to unlock earlier epochs in the Universe, probe the mechanisms
of galaxy formation, and gain insight into the evolution of stellar systems.

Models for the chemical evolution of galaxies need to account for the
collapse of gas and metals into stars, the synthesis of new elements within 
these stars, and the subsequent release of metal-enriched gas as stars lose 
mass and die. An additional feature of most models is the ongoing accretion of
gas from outside the system.  The most sophisticated models also incorporate a
self-consistent treatment of the system's dynamics - both collisionless and
dissipative components - either under idealised (semi-cosmological) conditions
or within a full cosmological framework.  Coupling GCE codes to a
spectrophotometric evolution package further ensures that the models are
constantly tested against real-world observational constraints.

Semi-analytic homogeneous models make simplifying assumptions that enable
the \emph{mean} trends of galactic systems to be calculated by numerically
solving a set of equations governing the formation, destruction, and
distribution of the elements as they cycle through gas and stars. One
strength of these models is that they typically have the fewest number of
free-parameters, making convergence to a \emph{unique} solution more likely.
A weakness of homogeneous models is the inherent assumption that stellar
ejecta from dying stars
is instantly mixed back into the ambient interstellar medium (ISM).
Inhomogeneous GCE models relax this so-called ``instantaneous mixing
approximation'' in a semi-analytical manner, allowing
consideration of observed trends in \emph{dispersion} in various 
galactic observables.  The self-consistent treatment of not only GCE, but the
dynamics of a galaxy's gas, stars, and dark matter, remains the purview of 
chemodynamical codes.  Each of the above are complementary tools for
deconstructing the formation and evolution of systems such as our Milky Way:
semi-analytical models can cover a range of parameter space that a
chemodynamical code cannot, due to the many orders of magnitude difference in
the respective computational demands, while the latter afford a coupling of the
dynamics of the system to that of the GCE, in a manner not otherwise available.

In the review which follows, we present a biased overview of
contemporary research in the field of Galactic Chemical Evolution. For
seminal reviews tracing the development and principles of this topic,
the reader is referred to Tinsley (1980), Matteucci (2001), and
references therein. Section~2 summarises the most popular formation
and evolutionary scenarios and describes the relationships between
different components of the Galaxy. Principles of homogeneous and
inhomogeneous semi-analytical models are presented in \S~3 and
\S~4, respectively, while in \S~5, the state-of-the-art in
three-dimensional cosmological chemodynamical codes is reviewed.
Coupling the GCE predictions from both of these approaches to galaxy
evolution to the colour and luminosity information provided by
spectrophotometric codes is discussed in \S~6.  Potential future areas
of interest to the field are itemised in \S~7.

\section{Structure and Formation of the Galaxy}

%Gilmore, Wyse \& Kuijken (1989) present a particularly comprehensive
%review of the structure of the Milky Way.

The thin disk and bright inner bulge are the brightest components of
the Milky Way. They are housed within a much more extended and diffuse
spheroidal stellar halo -- as distinct from the dark matter halo. In
addition, a fourth main component -- the thick disk -- was identified
by Gilmore \& Reid (1983). The latter found that star counts obtained by an
extensive UK Schmidt photometric survey were well-fitted by a
multi-component stellar population model consisting of 1) a thin
disk, 2) thick disk, and 3) halo. This three-component model breaks
down within the central kpc of the Galaxy where a dense metal-rich
bulge begins to dominate (Gilmore, Wyse \& Kuijken 1989). The thick
disk component has an exponential scale height of 1350 pc -- about
four times greater than the scale height of the thin disk -- and
comprises $\sim$2\% of the nearby stars. Photometric studies of
external galaxies had already established that thick disks are common
to spirals (van der Kruit \& Searle 1982). A popular explanation for
the presence of thick disks is that mergers with smaller satellites
during early times heated the thin disk (Wyse \& Gilmore 1993).

In order to distinguish individual populations of stars, one wishes to
know not just their spatial distribution, but their
kinematics, chemical abundances, and ages. A complete dataset of this
information should enable one to reconstruct the formation and
evolution of the Milky Way. A wealth of past, present, and future surveys
and satellite missions (e.g. 
HIPPARCOS\footnote{\tt http://astro.estec.esa.nl/Hipparcos/}, 
RAVE,\footnote{\tt http://astronomy.swin.edu.au/RAVE/} 
GAIA\footnote{\tt http://astro.estec.esa.nl/GAIA/}) offer (or will offer)
the opportunity to determine 
the order in which the Galactic components formed, whether they
evolved independently of one another, and how important merging has
been in assembling the Milky Way. The mean age and metallicity of the
halo, thick disk, thin disk, and bulge are shown in
Table~1.

\begin{table}
\caption{Milky Way Properties}
\centering
\begin{tabular}{l|cccc}
\hline
  & {\scriptsize mean} & {\scriptsize mean} & {\tiny Scaleheight} & {\tiny Scalelength }  \\
  & {\scriptsize Age (Gyr)$^a$ } & {\scriptsize [Fe/H]$^a$} & {\scriptsize  (kpc)} & {\scriptsize  (kpc)}  \\
\hline
{\scriptsize Halo}        & 14  & $-$1.78   &
\multicolumn{2}{c}{{\scriptsize Effective Radius}$\sim$2.7$^b$ }  \\
{\scriptsize Thick disk}  & 11  & $-$0.78   & $\sim$ 0.75$^c$ &3.5$^c$  \\
{\scriptsize Thin disk}   & 5-7  & $-$0.14  & $\sim$ 0.33$^c$ &2.25$^c$  \\
{\scriptsize Bulge}       & 10  & 0       &  \multicolumn{2}{c}{{\scriptsize Effective Radius}$\sim$1.2$^d$ }  \\
\hline
\end{tabular}
\begin{list}{}
\item {\scriptsize $^a$Robin et~al. (2003)\\ $^b$de Vaucouleurs profile -
Buser et~al. (1998)\\ $^c$Chen et~al. (2001)\\ $^d$de Vaucouleurs profile -
Yoshii \& Rodgers (1989)\\   }
\end{list}
\end{table}

Traditionally two scenarios have competed to explain the formation of
the Milky Way:
\begin{description}{}
\item  1) The first scenario, proposed by Eggen, Lynden-Bell \&
Sandage (1962), describes the rapid monolithic collapse of a
protogalactic gas cloud to form the halo. The Galactic disk would have
subsequently formed as the residual gas disipationally collapsed. This
would naturally give rise to two population of stars: an older, more
metal-poor group found in the halo; and a younger, more metal-rich group
orbiting closer to the Galactic mid-plane.
\item  2) Searle \& Zinn (1978) offered an alternative to the
monolithic collapse picture, proposing that the Galaxy was constructed
from smaller cloud fragments, in which stars may have already started
forming.
\end{description}
The Galaxy's true formation history is likely to lie somewhere between
the two extremes of primordial collapse and hierarchical
formation. Chemical properties of stars provide important clues into
disentangling the
puzzle of the Galaxy's formation. The relative abundances of certain
elemental species act as ``cosmic clocks'', by which the formation
timescales of various stellar populations can be determined. A popular
cosmic clock is the ratio of an element like oxygen, that is born
mostly in massive very short-lived stars, and an element like iron,
whose creation is linked to lower mass longer-lived progenitors
(Gilmore, Wyse \& Kuijken 1989).  

\section{Homogeneous Models}

Homogeneous GCE models have traditionally formed the cornerstone of
this field and consequently have a rich literature to draw upon.  The
basic ingredients, observational constraints, and several weaknesses
are highlighted below, although it should be emphasised that much of
this discussion pertains also to the inhomogeneous and chemodynamical
models described in \S~4 and \S~5 (as many of the ingredients are
common to all techniques).

\subsection{Basic Ingredients}

The main ingredients of homogeneous GCE models are outlined below and we
discuss their relationship to one another through the basic set of
chemical evolution equations.

\begin{itemize}

\item \textbf{Stellar Yields and Lifetimes:} Almost all elements heavier
than helium originate from stars. Stars enrich the ISM with their own unique
pattern of elements depending on their mass and initial metallicity. The
predicted stellar yields consequently form the backbone of the study of
GCE. The grids of yields utilised in the models are the outcome of
computations of stellar evolution and vast networks of nuclear reactions.
For the purposes of chemical evolution, stars are often divided into three
categories:

\begin{description}
  
\item \emph{Massive stars} ($m$~$\simgt$~10~M$_{\odot}$) evolve
  quickly because their enormous gravitational potential accelerates
  the nucleosynthesis process.  Their death is marked by a violent
  supernova (SN), leaving behind a neutron star or black hole.
  Although massive stars are much rarer than their lower mass
  counterparts, they are the main source of most of the heavy elements
  (i.e.  metals) in the Galaxy. Figure~1 shows the production factors
  from massive stars predicted by the detailed nucleosynthesis
  calculations of Woosley \& Weaver (1995, \emph{left panel}) and the
  FRANEC code 
  (kindly provided by A. Chieffi 2003, priv comm; FRANEC is described in 
  Chieffi \& Limongi 2002, \emph{right panel}).\footnote{A generic comparison
  of massive star yields can be found in Gibson, Loewenstein \& Mushotzky
  (1997).} The dotted line at
  [X/O]~=~0 indicates the solar elemental abundance pattern relative
  to O. The solar abundance pattern of most metals is adequately reproduced
  by massive stars but C, N and the iron-peak elements require
  additional production sites.

\item \emph{Intermediate- and low-mass stars (ILMS)} live longer than their
massive star counterparts, due
to their lower density. They greatly outnumber the heavier stars
but do not produce significant quantities of many elements besides helium,
carbon, nitrogen,
and certain isotopes, which are created through hydrostatic burning
and expelled in stellar winds and planetary nebulae (e.g. van den Hoek \&
Groenewegen 1997). Very low mass stars
($m$~$\simlt 1$~M$_{\odot}$) have lifetimes comparable to the age of the
Galaxy and therefore serve to lock up the gas supply.

\item \emph{ILMS in binary systems} may culminate in powerful supernovae
explosions classed as Type~Ia. The exact physical mechanisms 
behind SNe~Ia are still an open
question but one popular theory holds that the mass lost by a binary star as
it evolves is accreted by its smaller white dwarf (WD) companion until the
WD can no longer be sustained by electron degenerate pressure. Then the
entire mass of the WD is ejected in a violent explosion that converts much
of the stellar material into iron (e.g. Iwamoto et~al. 1999).

\begin{figure*}[t]
\begin{center}\hbox{
\psfig{file=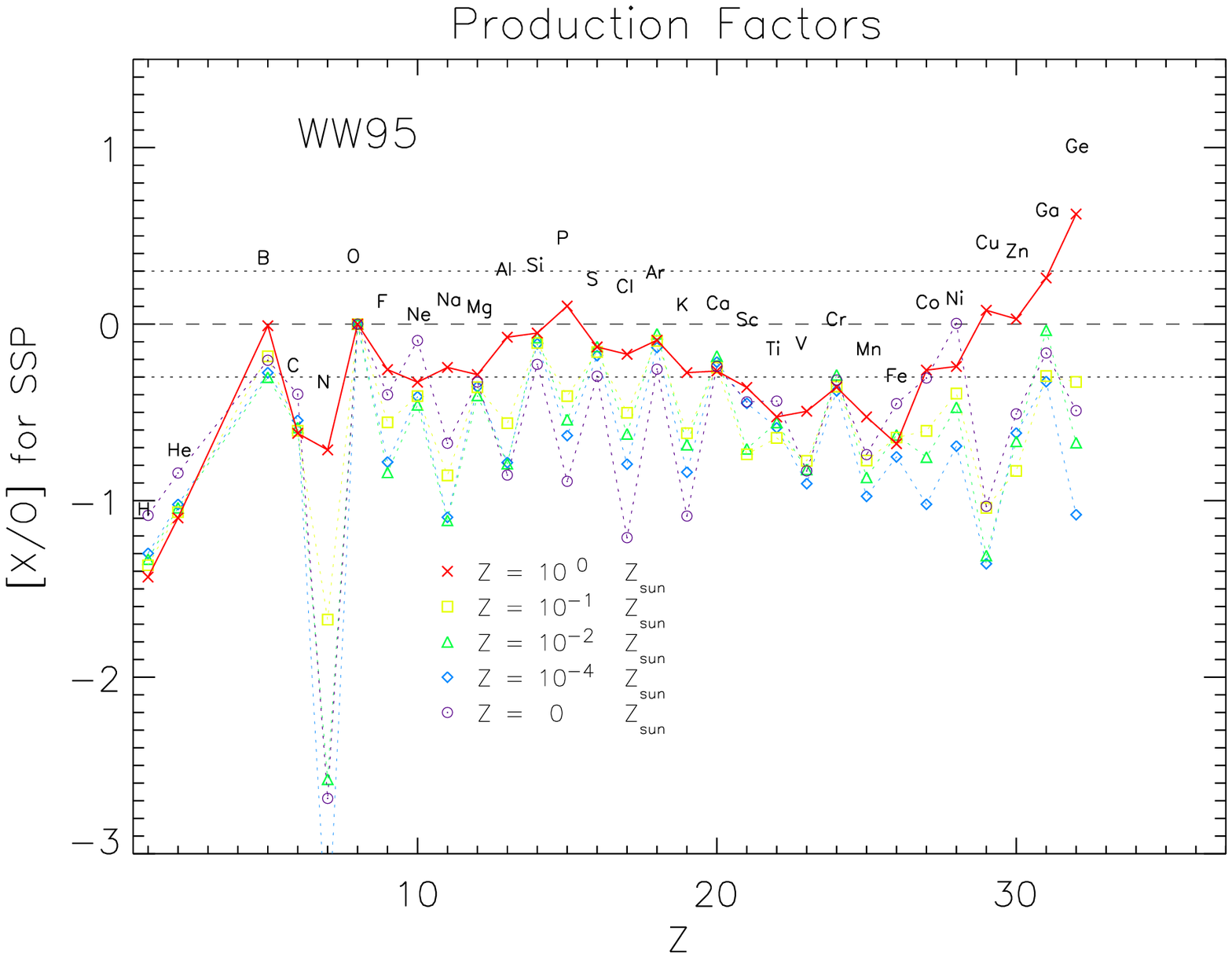,width=8.5cm}
\psfig{file=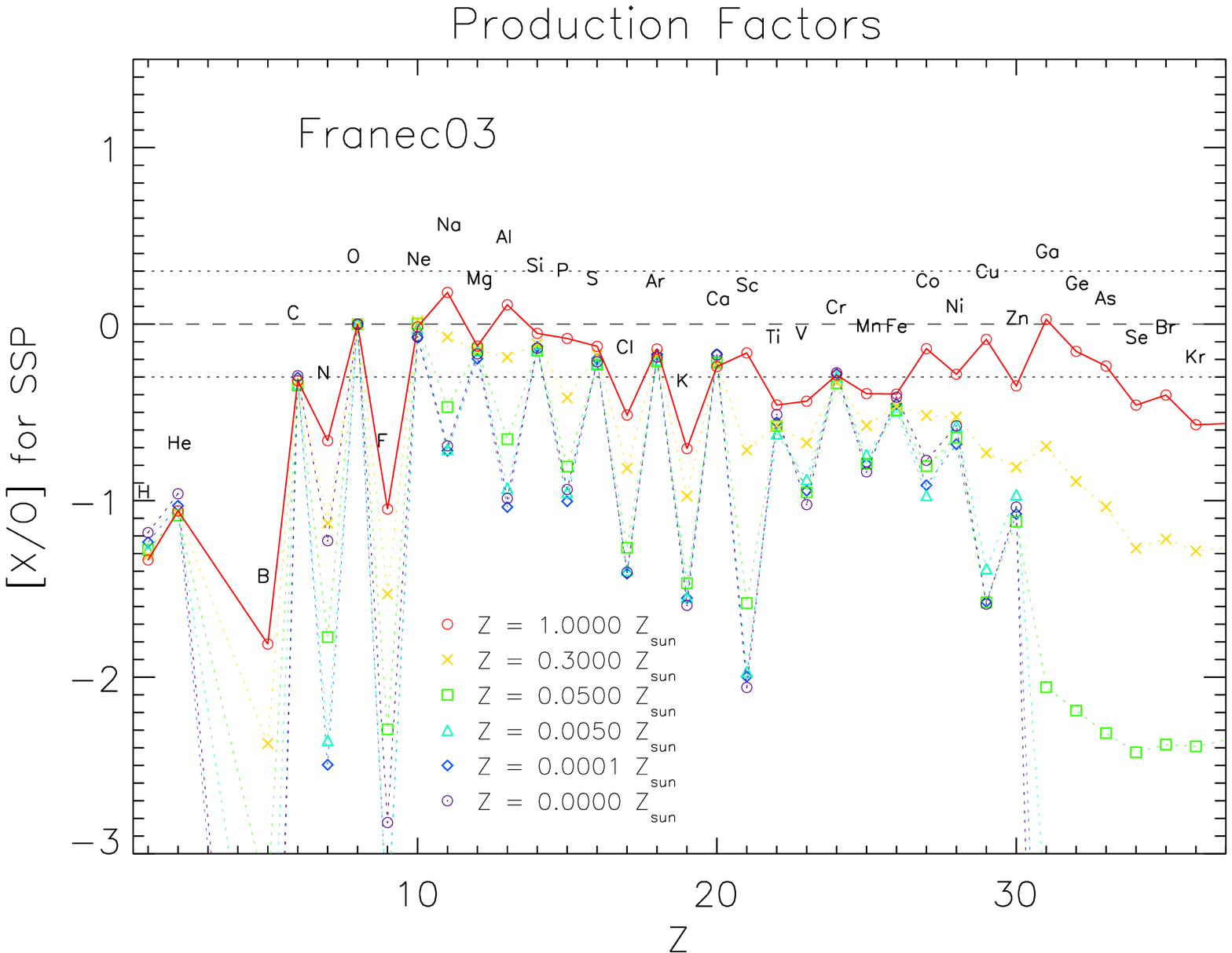,width=8.5cm}}
\caption{Production factors relative to O on a solar
  logarithmic scale from a single generation of massive stars using
  the metallicity-dependent yields of Woosley \& Weaver 1995
  (\emph{left panel}) and those of FRANEC 2003 (\emph{right panel}).
  The latter were kindly provided by A. Chieffi (2003, priv comm).
  Yields were integrated over a Salpeter (1955) IMF from 12-40
  M$_{\odot}$. The dashed line indicates the solar values (where
  log(N$_{\mathrm{O}}$/N$_{\mathrm{H}}$)$_{\odot}$ + 12 = 8.73
  (Holweger 2001)) and dotted lines indicate deviations from
  scaled solar by a factor of two. For both sets of yields C, N, and
  some of the iron-peak elements are subsolar because they require
  additional sources such as lower mass stars and Type~Ia SNe. The
  strength of the ``odd-even'' effect increases with decreasing
  metallicity in both cases, however the effect is more pronounced for
  FRANEC 2003.  }
\end{center}
\end{figure*}

\end{description}

\item \textbf{Initial Mass Function (IMF):} The precise form of the IMF
dictates the number of stars born in a given mass interval in each
generation of stars. This in turn sets the rate at which different
elements are released into the ISM, thus influencing both the
\emph{relative} and \emph{absolute} elemental abundances. Most IMFs in the
literature consist of simple single- or multi-component power law
specified over a mass range from $m \sim 0.1$~M$_{\odot}$ to an upper mass
limit typically between 40--100~M$_{\odot}$ (e.g. Salpeter 1955; Scalo
1986; Kroupa, Tout \& Gilmore 1993).

\item \textbf{Star Formation Rate (SFR):} While laws of star formation can
be calculated from first principles, chemical evolution models invariably
use a functional form that has been derived empirically. An ample supply
of gas is the first condition needed for star formation, so it is not
surprising that one of the simplest laws has SFR $\propto \sigma_{gas}^k$,
where $\sigma_{gas}$ is the surface density of gas and the exponent $k$
may range from 1--2 (Schmidt 1959). Other star formation laws presume that
factors such as total mass density and/or Galactocentric radius play a
role. Dopita \& Ryder (1994) found that a law given by SFR $\propto
\sigma_{gas}^{5/3} \sigma_{total}^{1/3}$ satisfactorily describes the
correlation between H$\alpha$ emission and I-band surface brightness in
spiral galaxies.

\item \textbf{Gas Flows:} In the simplest scenario, our model Galaxy in
each radial bin can be considered a closed-box consisting of
primordial gas from which stars are born according to the chosen SF
and IMF prescriptions. In the classic closed-box model (e.g. Pagel \&
Patchett 1975) there is no gas loss or gain; at time~$=$~0 all
Galactic matter is present as primordial gas from which stars
immediately form. This type of model, characterised by an intense
period of early star formation, provides a reasonable account of the
formation of the halo and bulge of the Milky Way. When applied
to the Galactic disk, however, the basic closed-box model leads to an
excess of metal-poor stars with respect to the observed metallicity
distribution of nearby long-lived stars: the so-called ``G-dwarf
problem'' (Pagel \& Patchett 1975). More realistic models overcome
this problem by allowing the Galactic disk to form via continual
accretion of gas. The infall rate as a fucntion of radius $r$ and time
$t$ often takes the form:

\begin{equation}
\displaystyle\frac{d}{dt} \sigma_{gas}(r,t) = A(r)e^{-t/\tau(r)},
\end{equation}

where $\tau(r)$ is the exponential infall timescale and the
coefficient $A(r)$ must satisfy the constraint that
$\int^{t_{now}}_{0} A(r)e^{-t/\tau(r)} = \sigma_{gas}(r)$, where
$\sigma_{gas}(r)$ is the present-day surface density profile.  Many
elliptical and dwarf galaxies are expected to have undergone periods
in which the energy from supernova explosions exceeds the
gravitational energy of the system, causing the interstellar gas to be
expelled in a galactic wind (Larson 1974; Dekel \& Silk 1986; 
Matteucci \& Tornamb\'e 1987; Gibson 1997;
Recchi, Matteucci, \& D'Ercole 2001). However, the outflow of gas is
not expected to feature in the history of most spiral galaxies and is
usually neglected in the models. Radial flows of gas have also been
investigated within the homogeneous GCE paradigm (e.g. Portinari \&
Chiosi 2000).

\item \textbf{Galactic Components:} The infall of gas described above
might occur in multiple episodes that correspond to the formation of the
individual halo, bulge, thick disk and thin disk components. Recent works
have adopted different premises regarding the extent to which these
components interact with each other through gas exchange. For instance,
enriched halo gas might be funneled into the bulge or else
might collapse further to form the disk. Alternatively, the
halo and thin disk might evolve coevally yet independently. The
answer to these questions is partly clouded by challenges in
distinguishing the various stellar populations. The properties that we
wish to measure in each population, such as metallicity and kinematics,
are also the properties used to define the populations.

\end{itemize}

\subsubsection{The Equations}

The chemical composition of the interstellar medium as a function of
time and Galactocentric radius is described by equations that balance:
1) processes that \emph{deplete} chemical elements from the
interstellar medium, namely star formation and perhaps Galactic winds;
and 2) processes that \emph{replenish} the ISM, such as stellar winds,
SNe, and infalling gas. The GCE equations can be solved analytically
given the assumption that stars release their ejecta instantaneously
at the time of their birth. This approximation is reasonable only for
very massive and short-lived stars and it precludes one from
reproducing the relative trends of elements that are restored to the
ISM by different mass stars on different characteristic
timescales. For GCE models to have the power to interpret the
variation of diagnostic abundance ratios, the ``instantaneous
recycling approximation'' must be relaxed. There is another
assumption, however, that few GCE models relax: namely, that the
stellar yields are instantaneously mixed into the ISM.  For this
reason, homogeneous GCE models are not the ideal tools for studying
the Galaxy in its infancy, when chemical abundances in the ISM could
be traced to individual supernovae (or for present-day observables
that originate from that era - e.g. the Galactic halo).  We return to
this assumption in \S~4.

Standard models of spiral galaxy chemical evolution assume azimuthal
symmetry and collapse the object into a flat disk such that radius is the
only spatial variable. Defining $\sigma_i(r,t)$ as the mass surface
density of species $i$ at radius $r$ and time $t$, then the rate of change
of $\sigma_i(r,t)$ is given by:

\begin{eqnarray}
\displaystyle\frac{d}{dt} \sigma_{i} (r,t) & = & 
 \displaystyle\int^{m_{up}}_{m_{low}}\psi(r,t-\tau_{m})\,Y_i(m,Z_{t-\tau_{m}})
 \,\frac{\displaystyle\phi(m)}{m} \,\; dm \nonumber \\
  & & + \displaystyle\frac{d}{dt} \sigma_i(r,t)_{infall} \nonumber \\
  & & -X_i(r,t)\,\psi(r,t) \nonumber \\
  & & - \displaystyle\frac{d}{dt}\sigma_i(r,t)_{outflow}
\end{eqnarray}

\noindent
where the four terms on the right-hand side of equation~(2) correspond to the
stellar ejecta, gas infall, star formation, and gas outflow rates, 
respectively.
$\psi$ is the SFR, $Y_i(m,Z_{t-\tau_m})$ is the stellar yield of $i$
(in mass units) from a star of mass $m$ and metallicity $Z_{t-\tau_m}$,
$\phi(m)$ is the IMF by mass, and $X_i$ is the mass fraction of element $i$. By
definition, the sum of $X_i$ over all $i$ is unity, and the total
surface mass density is identical to the integral over the infall and
outflow rates. $m_{low}$ and $m_{up}$ are the lower and upper stellar mass
limits, respectively,
and $\tau_{m}$ is the lifetime of a star of mass $m$. In practice,
the first term is split into three equations that deal separately with
ILMS, Type~Ia SNe progenitors, and massive stars (see Greggio \&
Renzini 1983 and Matteucci \& Greggio 1986 for details).

\subsection{Observational Constraints}

The most thoroughly observed and best understood galaxy is the Milky Way,
and in particular, the ``local'' solar neighbourhood. So extensive is the
Milky Way dataset, many model ingredients can be well-constrained
empirically. Thus our own Galaxy is often the gauge by which chemical
evolution models are calibrated. Indeed, studies of the cosmic evolution
of disk galaxies often adopt scaling laws based on the Milky Way (e.g.
Boissier \& Prantzos 2000). The extent to which the Milky Way's IMF,
star formation law, and nucleosynthetic behaviour can be applied
to other types of objects at earlier epochs depends on how universal these
prescriptions are and on whether we live in a prototypical galaxy. A
minimal set of observational constraints for GCE models is described
below.

\begin{itemize}

\item \textbf{Solar abundance pattern:} Any chemical evolution model
should be able to reproduce the solar abundance pattern, i.e. the pattern
in the ISM 4.5~Gyr ago at the radius where the Sun was born. The Sun is
the single star with the most complete set of abundance measurements. For
consistency, one can also compare the chemical composition of solar system
meteorites with estimates based on stellar spectral lines. As discussed
above, the predicted solar enrichment pattern is chiefly controlled by the
yields released in 1) Type~II SN explosions of massive stars, 2) planetary
nebulae and stellar winds of ILMS,
and 3) Type~Ia SN explosions of binary systems of ILMS. In addition,
the predictions are sensitive to the stellar mass distribution (i.e. the
IMF) and the SFR, since these set the relative
contribution and enrichment timescale from different types of stars. A seminal
attempt to simulate the evolution of all elements up to and including zinc was
made by Timmes, Woosley \& Weaver (1995, hereafter TWW95) using the 
comprehensive
grid of mass- and metallicity-dependent yields calculated by Woosley \&
Weaver (1995). Impressively, TWW95 reproduced the abundances of most
isotopes in the Sun to within a factor of two. 

It is worth noting that even our understanding of 
the solar abundance pattern still has the power to surprise us.  As recently as
a decade ago, the solar oxygen abundance was assumed to be 
log(N$_{\mathrm{O}}$/N$_{\mathrm{H}}$)$_{\odot}$ + 12 = 8.93
(Anders \& Grevesse 1989); it has become clear recently though
that accounting for solar
granulation and non-LTE effects properly leads to a \emph{significantly}
radical downward revision in the Sun's oxygen abundance by almost a factor of
two (to log(N$_{\mathrm{O}}$/N$_{\mathrm{H}}$)$_{\odot}$ + 12 = 8.69 -
Allende Prieto, Lambert \& Asplund 2001).  Such a shift
partially resolves the long-standing dichotomy between the Sun's oxygen
abundance and that of the local ISM.

\item \textbf{G-dwarf distribution:} A much stronger constraint on GCE
models is the distribution of stars as a function of metallicity, since
this represents the convolution of the age-metallicity relationship and
the star formation history. In order to probe the early Galaxy, one needs
a sample of low mass stars such as G- or K-dwarfs whose main-sequence
lifetimes are comparable to, or older than, the age of the Universe. GCE
models have demonstrated that the paucity of low-metallicity dwarf stars
can be explained if the Galactic halo formed first on a rapid timescale,
followed by a slow build-up of the thin disk (e.g. Chiappini et~al. 1997;
Alib{\' e}s et~al. 2001; Fenner \& Gibson 2003). The excess of
metal-poor stars predicted by simple closed-box models can also be 
avoided by assuming
prompt initial enrichment, perhaps by a first generation of extremely
massive Population~III stars. If this were the case, then the abundance
pattern of Pop~III ejecta should be evident in the lowest metallicity
stars. It has also been suggested that there are no very metal-poor
stars left because they all had relatively short lifetimes due to low
metallicity environments favouring the formation of higher mass stars
(e.g. Nakamura \& Umemura 2001). A further consideration is that an 
initially pristine
zero-metallicity star might have had its surface layers
polluted by the accretion of metals
from the ISM over the past $\sim$12~Gyr (e.g. Shigeyama, Tsujimoto \& Yoshii
2003).

\item \textbf{Evolution of abundance ratios:} If elements X$_1$ and X$_2$ have
different origins and different characteristic timescales for release
into the ISM, then [X$_1$/X$_2$] vs [X$_2$/H] acts as a clock by which
chemical evolution can be measured (e.g. Wyse \& Gilmore 1988).  For
example, readily observable features of oxygen and iron in stellar
spectra has encouraged the wide use of [O/Fe] vs [Fe/H] to diagnose
the overall star formation history of galactic systems. As with most
heavy elements, oxygen is produced chiefly in massive and short-lived
stars.  Thus, oxygen enrichment immediately follows the onset of star
formation. In contrast, at least half of the iron in the Galaxy
probably originated from Type~Ia SNe (e.g. Alib\'es et~al.
2001), whose lower-mass and longer-lived progenitors introduce a time
delay for iron enrichment. The remaining iron comes largely from
Type~II SNe. The combination of high [Fe/H] and high [O/Fe] is
understood to arise in systems that formed stars so rapidly that high
metallicities were reached before SNe~Ia had a chance to lower the
[O/Fe] value (Smecker-Hane \& Wyse 1992). Similarly, one might
interpret low [Fe/H] and low [O/Fe] as a sign of a slowly evolving
system.

\item \textbf{SFR and SN rates:} The present-day star formation and
Type~II and Type~Ia SNe rates must be matched by a successful chemical
evolution model. However these are fairly weak constraints given that
we can only be reasonably certain about the current SFR and mean past
rate. Finer details of the Galactic star formation history are
difficult to recover and are quite uncertain. The most direct way to
infer the SFH is by determining the age distribution of stars; a
method that relies upon unreliable stellar ages and assumptions about
the IMF, stellar evolution, scale height corrections, and stellar
kinematics (Rocha-Pinto et~al. 2000a). This technique is also somewhat
circular, in that a SFH must have been assumed in order to derive the
IMF.

\item \textbf{Age-metallicity relationship:} This is an important
constraint, but again, a weak one given that the scatter in the
observations (e.g. Ibukiyama \& Arimoto 2002) can accommodate most
model predictions. Moreover, the very existence of an AMR, which had
been well established by earlier studies (e.g Twarog 1980; Edvardsson et~al. 1993; 
Rocha-Pinto et~al. 2000b), has recently been challenged by investigations
demonstrating large intrinsic scatter and no significant trend of 
metallicity with age (e.g. Feltzing, Holmberg \& Hurley 2001).

\item \textbf{Gas and Abundance gradients:} It has long been known
  that the Milky Way is more metal-rich toward its centre and more
  metal-poor at large Galactocentric distance (Tinsley 1980, and
  references therein).  Using the oxygen abundance observed in
  H\,{\small II} regions and OB stars to trace metallicity, a
  metallicity gradient of $-$0.07~dex~kpc$^{-1}$ has been established
  (e.g. Smartt \& Rolleston 1997). The abundance of metals in a region
  of gas is particularly sensitive to the balance between the star
  formation and gas accretion rates. Therefore the predicted
  metallicity gradient of the Galactic disk depends strongly on how
  the star formation prescription and gas infall rate are assumed to
  vary with radius. Good fits to the data are obtained by
  ``inside-out'' formation scenarios, whereby the innermost disk is
  built-up on the shortest timescale (e.g. Larson 1976; Chiappini,
  Matteucci \& Gratton 1997). Portinari \& Chiosi (1999) showed that a
  SFR such as the Schmidt (1959) law, which varies only with the gas
  surface density, produces a radial abundance profile that is too
  flat unless one assumes an unreasonably large variation in formation
  timescale from the inner to outer disk such that the far disk would
  currently be accreting at much higher rates than observed. The
  Dopita \& Ryder (1994) law, with a mild dependence on \emph{total}
  mass surface density, yields a better fit to the metallicity
  gradient (Portinari \& Chiosi 1999).  The theoretical metallicity
  distribution of long-lived stars as a function of Galactocentric
  radius is shown in Figure~2, assuming a Dopita \& Ryder (1994) star
  formation law.

\begin{figure}[ht]
\begin{center}
\psfig{file=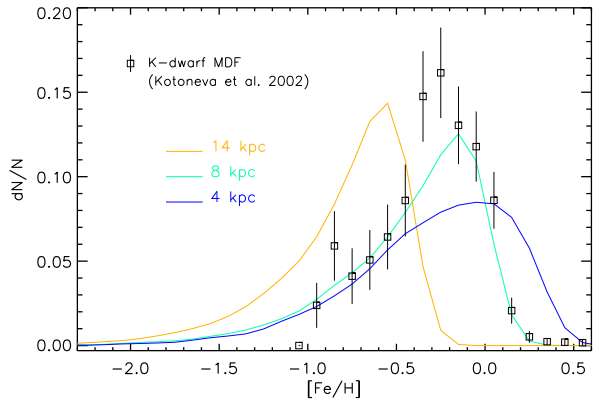,width=8.5cm}
\caption{The metallicity distribution function (MDF) of long-lived stars predicted by
  a single infall model at three different Galactocentric radii; 4, 8,
  and 14 kpc.  Open squares with error bars correspond to the MDF of
  nearby K-dwarfs (i.e. at a radius $\sim$ 8-8.5 kpc) observed by
  Kotoneva et~al. (2002). The theoretical MDF has been convolved with a
  Gaussian of dispersion $\sigma$~=~0.1~dex in [Fe/H] to simulate the
  empirical uncertainties.  }
\end{center}
\end{figure}

\item \textbf{Isotopic abundances:} Traditionally, chemical evolution
studies have been concerned with monitoring the \emph{total} abundance (or
dominant isotope) of specific elements in order to unravel the Galaxy's
history. Recent advances in instrumentation have paved the way for
research into individual isotopes that provides new challenges for
nucleosynthesis theory. For instance, Type~II SNe models appeared
capable of explaining the magnesium isotopic ratios in intermediate to
solar metallicity stars, but the results from the solar neighbourhood
model shown in Figure~3 reveal an underproduction of $^{26}$Mg/$^{24}$Mg
at low metallicites (\emph{dotted line}) with respect to the latest data.
The missing piece of the puzzle may be low metallicity intermediate mass
stars on the
asymptotic giant branch (AGB), whose helium-shells may be hot enough to
generate $^{25}$Mg and $^{26}$Mg by triggering $\alpha$-capture onto
$^{22}$Ne (Karakas \& Lattanzio 2003).  As Fenner et~al. (2003) have
demonstrated for the first time, the data at [Fe/H]~$\simlt$~$-1$ is
much better matched after incorporating the Karakas \& Lattanzio AGB
nucleosynthesis calculations in a chemical evolution model (\emph{solid
line}). Such detections of isotopic ratios in field stars and globular
clusters may reveal a great deal about the relative role of different
types of stars in various environments.

\begin{figure}[ht]
\begin{center}
\psfig{file=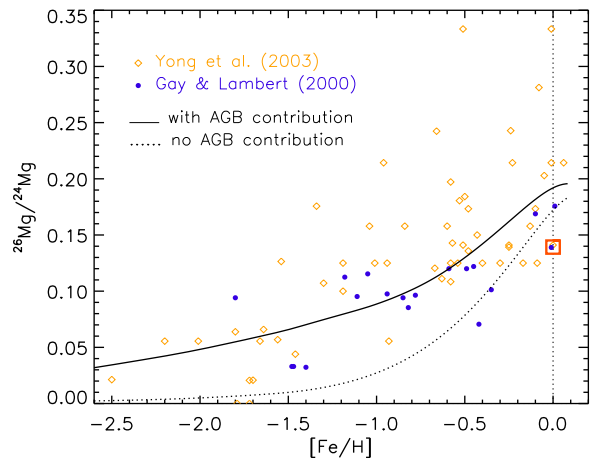,width=8.5cm}
\caption{The predicted metallicity evolution of magnesium isotopic ratio
$^{26}$Mg/$^{24}$Mg for the solar neighbourhood both with (\emph{solid
curve}) and without (\emph{dotted curve}) a contribution from
intermediate-mass AGB stars (from Fenner et~al. 2003). 
Stellar data are from Gay \& Lambert (2000,
\emph{blue circles}) and from halo and thick disk stars from Yong et~al.
(2003,
\emph{orange diamonds}). The red square denotes the solar ratio.}
\end{center}
\end{figure}

\end{itemize}

\subsection{Uncertainties and Weaknesses}

\begin{itemize} 

\item \textbf{Iron-peak Yields:} Iron-peak elements are buried deep within
the cores of massive stars near the radius that separates the ejected
material from the remnant. The location of this so-called ``mass cut'' is
a free-parameter in stellar models, one which controls the relative abundances
of the iron-peak elements as well as the X$_i$/Fe ratio in the ejecta.
Abundances in very metal-poor stars can help constrain the choice of mass
cut, however in order to simultaneously eject iron-peak elements in the
correct proportions \emph{and} recover the high observed [$\alpha$/Fe]
ratios, models need to incorporate mixing and fallback (Umeda \& Nomoto
2002) or asymmetrical explosions.  Multi-dimensional simulations of
explosive nucleosynthesis may reveal more about these processes
(Travaglio, Kifonidis \& M\"uller 2003).

\item \textbf{Shape and evolution of the IMF:} A time-invariant IMF
  remains the best choice for modelling the general evolution of our
  own Galaxy (Chiappini, Matteucci \& Padoan 2000), but peculiar
  abundance patterns in extremely metal-poor stars support the notion
  that the first generation of stars was biased towards higher masses
  (Chieffi \& Limongi 2002). The upper limit of the stellar IMF,
  m$_{\mathrm{up}}$, is also uncertain and impacts upon the total
  amount of metals produced by each stellar generation. Figure~4
  illustrates the the sensitivity of metal-growth in the solar
  neighbourhood to the upper IMF mass. Increasing m$_{\mathrm{up}}$
  from 40~M$_{\odot}$ to 100~M$_{\odot}$ is expected to raise the
  metallicity by as much as 30\%. This is due to the steeply
  increasing yield of O (which is the most abundant metal in the
  interstellar gas) as a function of stellar mass.

\begin{figure}[ht]
\begin{center}
\psfig{file=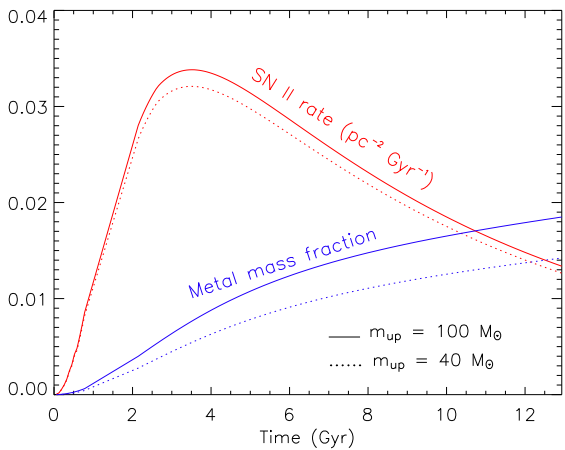,width=9.0cm}
\caption{Evolution of Type~II SN rate (\emph{red lines}) and
  metallicity (\emph{blue lines}) predicted by a single infall phase
  model of the solar neighbourhood for an IMF upper mass limit of 40
  M$_{\odot}$ (\emph{dotted lines}) and 100 M$_{\odot}$(\emph{solid
    lines}).  Metallicity is defined as the mass fraction of elements
  heavier than He in the interstellar medium (ISM). The SN~II rate is
  barely affected by changes in the upper limit to the (Kroupa et al.
  1993) IMF, however metallicity is about 30\% higher for the
  m$_{\mathrm{up}}$=100 M$_{\odot}$ case. This reflects the steep increase in
  the yield of O (which is the dominant element contributing to
  metallicity) as a function of initial stellar mass.}
\end{center}
\end{figure}

\item \textbf{Black hole mass limit:} Essentially a free parameter in Type~II
SNe models, the mass 
above which most of the stellar material collapses to
form a black hole must be addressed \it a posteriori\rm, in an empirical manner
(to recover the observable constraints alluded to earlier).
Black hole progenitors are expected to release most of
their oxygen, carbon, etc., via pre-SN stellar winds, but heavier elements such 
as iron
are expected to fall back onto the remnant. One may expect
the mass range corresponding to black hole collapse is sensitive to
metallicity (e.g. Maeder 1992).

\item \textbf{Limited dataset:} Many conclusions about cosmic chemical
evolution have been drawn in part from studies of the Milky Way, but the
Milky Way is only one object. How unique or typical the Milky Way is
amongst other galaxies is unknown and it can be dangerous to take the
Milky Way's evolutionary path as representative of most spirals. For
instance, a system such as M31, that resembles the Milky Way in terms of
size and morphology, shows evidence for dramatically different metallicity
distributions of its stellar populations (Worthey \& Espana 2003). As an
example of the perils of drawing conclusions based on the solar
neighbourhood, the almost constant ratio of Zn/Fe versus metallicity might
be taken as a sign that zinc, like iron, owes much of its production to
Type~Ia SNe (Matteucci et~al. 1993; Mishenina et~al. 2002). It is possible,
however, that zinc comes primarily from Type~II SNe, with a strongly
metallicity-dependent yield that mimics the time-delay associated with a
SN~Ia source (e.g. Timmes et~al. 1995). As a consequence, zinc would not
always vary in lockstep with iron but would depend on specific formation
histories. It is worthwhile to bear this in mind, as zinc is often used
as a proxy for iron in the high-redshift universe.

\end{itemize}

\section{Inhomogeneous Models}

The halo has special significance for the formation of the Galaxy. If halo
stars form a collisionless system, their orbits contain information about
the dynamics at the time of the formation. In addition, 
the abundances observed in low
mass stars reflect the chemical abundances and inhomogeneities during halo
formation. Element abundance ratios as a function of metallicity (generally)
show increasing scatter with decreasing metallicity,\footnote{As
Spite et~al. (2003) note though, this canonical ``wisdom'' needs some revision.}
whereas at higher
metallicities the scatter decreases to reach a mean element abundance
which corresponds to the ratio of the stellar yields integrated over the
IMF (the regime in which the homogeneous GCE models of \S~3 are most
applicable).

The enrichment of the halo mainly depends upon the number of SN
explosions and the manner in which the ejected gas is mixed with the
ambient ISM. If the mixing volume is sufficiently large, the
enrichment could be spatially homogeneous.  Conversely, should mixing
be inefficient, significant (localised) abundance inhomogeneities
could exist; in theory, gas in the vicinity of a SN might even bear
the chemical imprint of that \emph{single} event.  If the latter were
the case, second-generation, extremely metal-poor halo stars may show
an abundance pattern which matches the nucleosynthetic yields of a
single Pop~III SN.  An highly inhomogeneous halo would naturally
result in this case.  Simulating the temporal and spatial history of
chemical inhomogeneities in the halo of the Milky Way, through
semi-analytical GCE models, is an \emph{extremely} active field at the
present. Important earlier work in this field include investigations
by Malinie, Hartmann \& Mathews (1991), Malinie et~al. (1993),
Pilyugin \& Edmunds (1996), Copi (1997), van den Hoek \& de Jong
(1997), Ikuta \& Arimoto (1999), and Travaglio, Galli, \& Burkert
(2001). In what follows, we review three of the more recent approaches
to modelling inhomogeneous chemical evolution.

\subsection{Argast et~al. (2000,2002)}

Argast et~al. (2000,2002) evolve an $\sim$15\,kpc$^3$ region of mass
10$^8$\,M$_\odot$ with a spatial resolution of 50\,pc.
At each time step (1\,Myr), a certain number of cells are
randomly chosen, and each one can create a star with a probability
proportional to $\sigma_{gas}^2$ (akin to the Schmidt Law described in \S~3.1),
under the assumption of a Salpeter (1955) IMF.  Type~Ia SNe were not included in
their models, and the simulations were halted once [Fe/H] reached 1/10 solar.
Stars which form out of material enriched by a single SN will
inherit its abundances and show an elemental pattern which
reflects the progenitor mass. The Argast et~al. (2000) model has minimal
ISM mixing, as the expansion of the SN remnant is the only
dynamical process taken into account.

The results of their Inhomogeneous Galactic Chemical Evolution (IGCE) modeling
 - in relation to the trend of abundance ratio scatter as a function of [Fe/H]
 - are in good agreement with observations, except for [Cr/Fe], [Mn/Fe], and
[Ni/Fe].\footnote{The scatter in chromium and manganese does not increase
(observationally) with decreasing iron abundance, while the observed scatter
in nickel exceeds that predicted by their models.}
Argast et~al. conclude that for [Fe/H]$<-3$, the ISM is 
unmixed and dominated by local inhomogeneities polluted by individual
SN events. For [Fe/H]$>-2$, their model halo ISM reflects a true IMF-averaged
abundance pattern and is considered ``well-mixed''.  Argast et~al. do note
though that an individual SN event can still have an impact upon a well-mixed
ISM, leading to finite dispersions in abundance patterns at disk-like
metallicities.

\subsection{Oey (2000,2003)}

The initial condition of Oey's (2000) IGCE model is that of a
metal-free closed-box, in which the first generation of star forming
regions is randomly distributed, occupying a volume filling factor
$Q$. Each region behaves as a superbubble powered by its enclosed
Type~II SNe.  Oey then considers $n$ subsequent generations of star
formation, allowing the star forming regions to overlap.  The main
conclusion drawn by Oey (2000) is that the evolutionary state of a
system is characterized by the product $nQ$, with the relative filling
factor of contamination having the same importance as the number of
contamining generations. The high metallicity tail of the MDF may
provide a useful discriminant between the classical ``Simple Model'' and
the Simple Inhomogeneous Model proposed by Oey. The latter agrees with
both the Galactic halo and bulge MDFs by varying only this single
parameter $nQ$, with $Q$ and $n$ independent and roughly associated
with the global star formation efficiency and age.

The Simple Inhomogeneous Model assumes no large scale mixing beyond
the superbubble radii, with metals uniformly distributed within the
volumes of the hot superbubbles and cooling locally.
Once mixing is allowed, the metallicity would be reduced by dilution, and
an increase in $nQ$ would be required to attain a given present day
metallicity. Oey (2003) takes into account interstellar mixing processes in
ordinary multiphase ISM (mainly diffusion and turbulent 
mixing\footnote{Note though that
mixing between discrete enrichment regions is not allowed in Oey's (2003)
prescription.}), which does lead to lower metallicities within the superbubble.
As a result of this analysis, turbulent mixing would appear to
be more efficient than diffusion, but the lowered metallicity for parent
enrichment events requires more evolution (higher $nQ$) to match the
observed metallicities, and this in turn implies that the system's MDF drops
off too steeply to match the data in the high metallicity tail of the
Galactic halo MDF. Furthermore, the Pop~III stellar fraction is too high
compared to the observations, suggesting that a discrepancy remains
between the model and the observations. Future developments in the model are
eagerly anticipated.

\subsection{Tsujimoto et~al. (1999)}

Under the Tsujimoto et~al. (1999) formalism, halo star formation is confined to
separate clouds of mass $M_{c}$. Each
cloud is initially composed of Pop~III stars, with mass
fraction $x_{III}$, and gas that has yet to form stars. 
Subsequent generations of stars are assumed to form in SN remnant (SNR)
shells. The mass
fraction of each shell that turns into stars ($\epsilon$)
is assumed to be constant.
Heavy elements ejected from a SN event are
assumed to be trapped and well-mixed within the SNR shell. Some of these
elements go into stars forming of the next generation.
This process continues until remnants are no longer capable
of sweeping up sufficient gas to form shells.
The mass of a shell $M_{sh}(m,t) = M_{ej}(m) +
M_{sw}(m,t)$, where $M_{ej}(m)$ is the mass of the SN ejecta, and $M_{sw}(m,t)$
is the mass swept up by a shell (assumed to be 6.5$\times$10$^4$\,M$_\odot$
throughout their analysis).

The free parameters of the model are the mass fraction $x_{III}$ of metal free 
Pop~III stars initially formed in each cloud and the mass fraction $\epsilon$
of stars formed in each SNR. These values are chosen to reproduce the
observed [Fe/H] distribution function of halo field stars for
[Fe/H]$<$$-$1. If $x_{III}$ is too large, the total gas swept up
by the first SNRs exceeds the entire amount of available gas, and the star
formation stops at the first or second generation. If the process is to
continue as a sequence of SN-induced star formation, $x_{III}$ must be
$\simlt$$10^{-2}$ and $\epsilon$ confined to a narrow
range such that of order one massive star is born from each SNR.
If $\epsilon$ is too high, star formation soon stops with little
enrichment. If $\epsilon$ is too low, star formation will proceed until
most of the gas is used up, with an excess of enrichment. 

Tsujimoto et~al. (1999) conclude that the probability $p_{III}$ of observing a 
Pop~III star amongst the general background field of halo stars, under the
assumption that $M_c=10^6-10^7$\,M$_\odot$, should be $10^{-3} - 10^{-4}$ -
consistent with current observational limits (Beers 2000).  Their IGCE
model also naturally recovers the frequency distribution of stars in the
[Eu/Fe] $-$ [Fe/H] plane; [Eu/Fe] spans $>$2\,dex for [Fe/H]$<$$-$2, 
converging to a plateau by [Fe/H]$\approx$$-$1.  Further enhancements and
applications of the model are described in Suzuki \& Yoshii (2001),
Tsujimoto \& Shigeyama (2002), and Tsujimoto et~al. (2002).

\section{Chemodynamical Models}

GCE is intimately related to the Galactic star formation history, and star
formation is equally linked to the dynamical evolution of the Galaxy.
The self-consistent treatment of the chemical and dynamical evolution of 
a system has long been recognised as desirable, but the computationally
intensive nature of the simulations made this desire difficult to realise.
Advances in hardware and numerical methods over the past decade though
has finally allowed chemodynamical codes to realise their theoretical
promise (e.g. Larson 1976; Samland, Hensler \& Theis 1997; Carraro, Lia \&
Chiosi 1998; Nakasato \& Nomoto 2003; Brook et~al. 2003a). 
Recent three-dimensional chemodynamical evolution codes are now being applied 
routinely to study disk galaxies such as the Milky Way. 
For example, Steinmetz \& Muller (1995) succeeded in distinguishing 
the chemical properties between
halo, bulge, and disk stars (see also Bekki \& Chiba 2001). Raiteri,
Villata \& Navarro (1996) and Berczik (1999) took into account metal
enrichment by both Type~II \emph{and} Type~Ia SNe, reproducing
the correlation between [O/Fe] and [Fe/H] for
stars in the solar neighbourhood. Brook, Kawata \& Gibson (2003b) studied
the metal enrichment from intermediate mass stars and discussed how the
distribution of carbon, nitrogen, and oxygen abundances in the solar 
neighbourhood constrains the shape of the IMF (see also Gibson \&
Mould 1997).  Coupling chemodynamical models to spectrophotometric
codes (see \S~6) allows one to constrain, for example, the zero point of
the Tully-Fisher relation (e.g. Steinmetz \& Navarro 1999; 
Navarro \& Steinmetz 2000; Koda, Sofue \& Wada 2000; Abadi et~al. 2003a).

We now review the methodology of chemodynamical codes and their
application in modeling the evolution of the 
Milky Way.  The discussion is framed primarily within the context
of our N-body/Smoothed Particle 
Hydrodynamics software package {\tt GCD+} (Kawata 1999;
Kawata \& Gibson 2003a,b; Brook et~al. 2003a,b).

\subsection{Brief Introduction to {\tt GCD+}}

{\tt GCD+} was originally (Kawata 1999; Kawata \& Gibson 2003a) based upon
Katz et~al.'s (1996) {\tt TreeSPH} code, combining the 
tree algorithm (Barnes \& Hut 1986)
for the computation of the gravitational forces, with the smoothed
particle hydrodynamics (SPH) (Lucy 1977;
Gingold \& Monaghan 1977) approach to numerical hydrodynamics. The
dynamics of the dark matter (DM) and stars is calculated by the N-body
scheme, and the gas component is modeled using SPH. It is fully
Lagrangian, three-dimensional, and highly adaptive in space and time owing
to individual smoothing lengths and individual time steps. Moreover, it
includes self-consistently important physical processes,
such as self-gravity, hydrodynamics, radiative cooling,
star formation, SNe feedback, and metal enrichment.

Metallicity-dependent radiative cooling is taken into account, following
Sutherland \& Dopita (1993).  The
cooling rate for a gas with solar metallicity is larger than that for gas
of primordial composition by more than an order of magnitude. As such, the
cooling provided 
by metals should not be ignored in numerical simulations of galaxy
formation (Kay et~al. 2000).

{\tt GCD+} takes into account the energy feedback and metal
enrichment from both SNe~II and SNe~Ia, as well as enrichment from 
ILMS.  We
assume that each massive star ($\geq8\ {\rm M_{\rm \odot}}$) explodes as a
Type~II supernova. The SNe~Ia rates are calculated using the model
proposed by Kobayashi, Tsujimoto \& Nomoto (2000). The yields of SNe~II,
SNe~Ia, and ILMS
are taken from Woosley \& Weaver (1995), Iwamoto et~al. (1999), and 
van~den~Hoek \& Groenewegen (1997), respectively. The
mass, energy, and heavy elements are smoothed over the neighbouring gas
particles using the SPH smoothing algorithm. For example, when the $i$-th
star particle ejects a mass of $M_{\rm SN,{\it i}}$, a mass increment is
applied to the $j$-th neighbour gas particle, as such:
\begin{equation}
 \Delta { M_{\rm SN,{\it j}}}
  =  \frac{m_j}{\rho_{{\rm g},i}} { M_{\rm SN,{\it i}}}
  W(r_{ij}/h_{i}),
\end{equation}
\noindent
where
\begin{equation}
 \rho_{{\rm g},i} = \langle \rho_{\rm g}(\mbox{\boldmath $x$}_i) \rangle
 = \sum_{j \neq i} m_j W(r_{ij}/h_{i}),
\end{equation}
\noindent
and $W(x)$ is an SPH kernel (Kawata \& Gibson 2003a).  {\tt GCD+}
monitors the evolution of all relevant chemical elements self-consistently
with the dynamics of the gas, stars, and dark matter.  The photometric 
evolution of the stellar populations is also calculated simultaneously via
the Kodama \& Arimoto (1997) spectral synthesis package.

\begin{figure*}[ht]
\begin{center}
\psfig{file=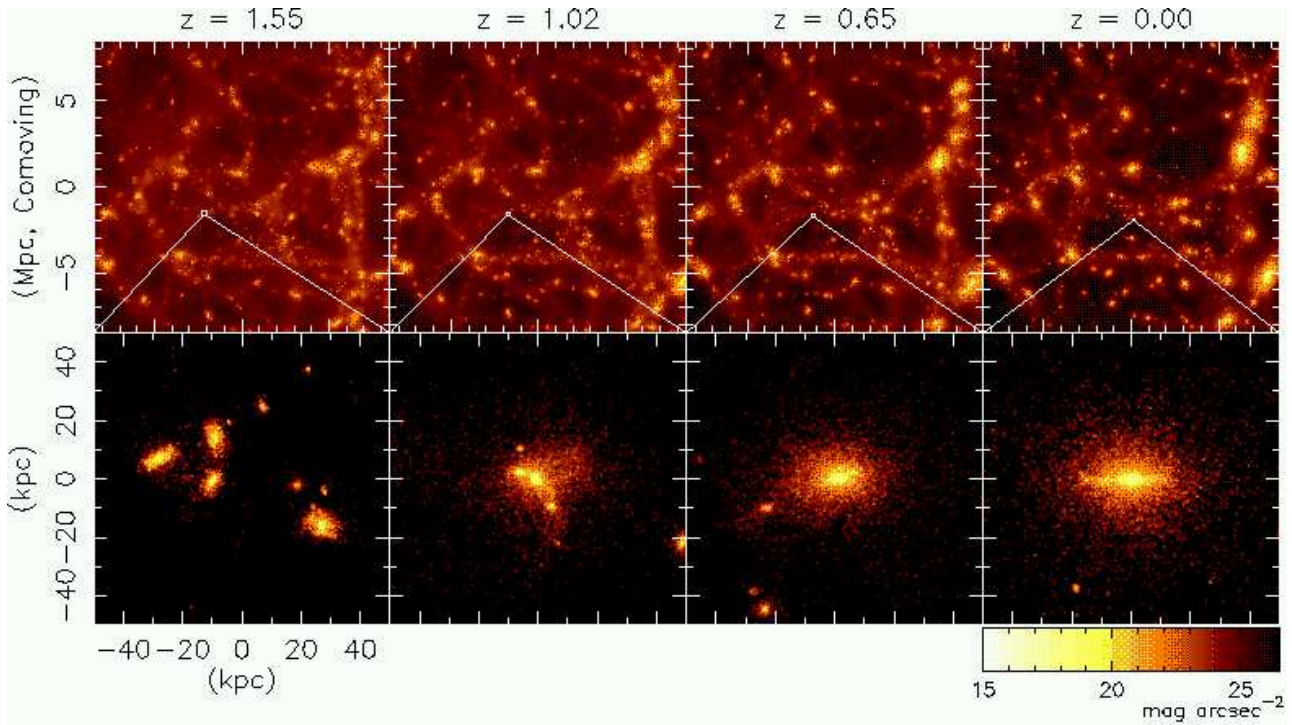,width=17.0cm} 
\caption{Dark matter density map of a portion of the 17~Mpc (comoving)  
simulation volume ({\it upper panels}), and predicted $J$-band (AB
magnitude) image
(physical scale) of the target galaxy ({\it lower panels}), over the
redshift range $z$=1.55 to $z$=0. The projection in the lower panels has
been chosen in order to view the target galaxy edge-on at $z$=0.
}
\end{center}
\end{figure*}

\begin{figure*}
\begin{center}
\psfig{file=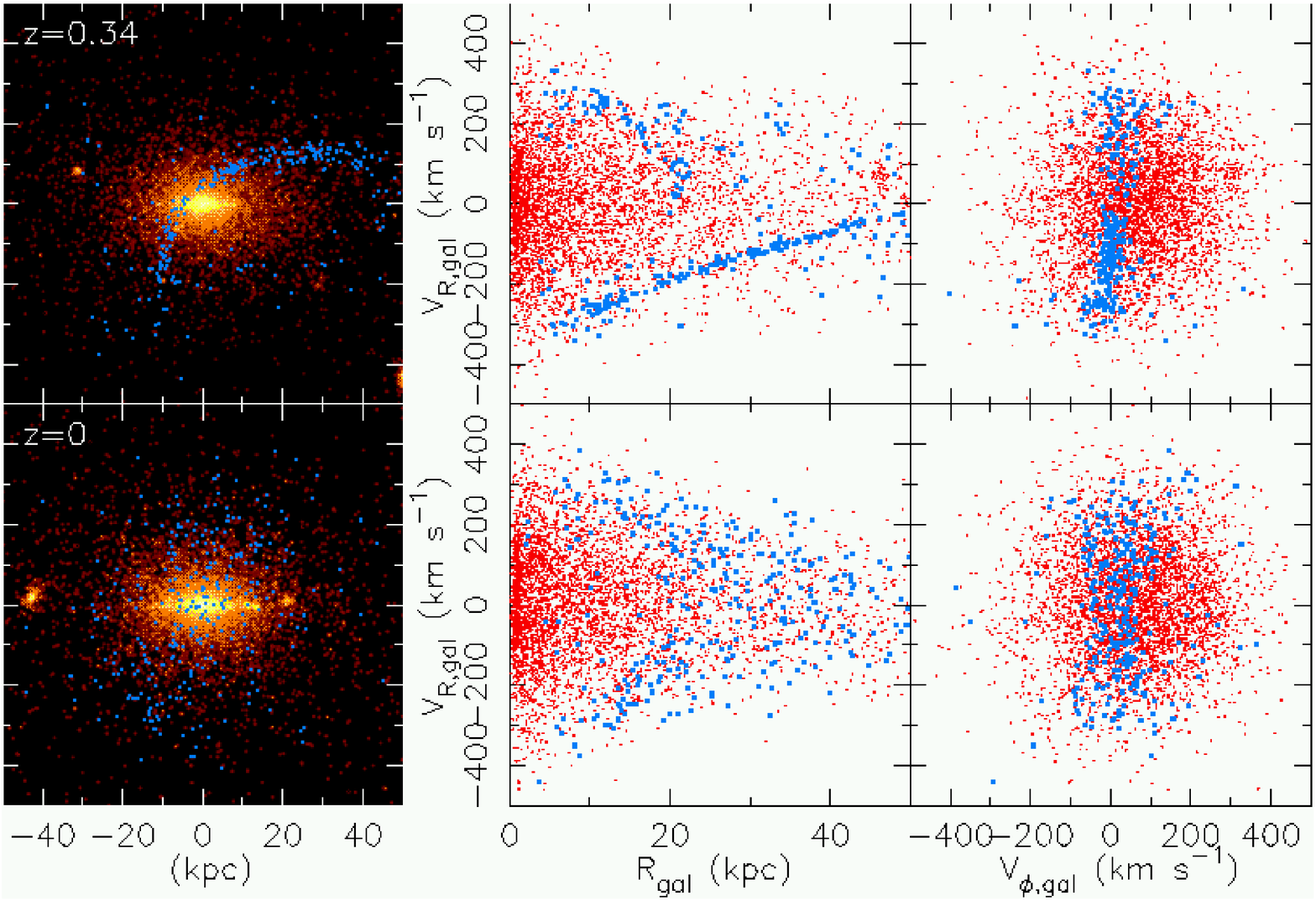,width=17.0cm} 
\caption{ 
Snapshots of the spatial and velocity distributions of
the member star particles (blue dots) of a satellite 
which accretes onto the central galaxy at $z\sim0.3$.
{\it Left panels} show the edge-on projection of the accreting satellite
as well as $J$-band (AB magnitude) images of the target galaxy 
(colour contour image, with levels as in Figure~5).
{\it Middle and right panels} show the radial velocity versus the
galactocentric radius (middle) and circular velocity (right)
of particles with galactocentric radius less than 50~kpc.
The red dots show star particles with [Fe/H]$<$$-$0.6, representative of the
population of halo stars in the central galaxy.
}
\end{center}
\end{figure*}

\subsection{Disk Galaxy Formation in $\Lambda$CDM Cosmologies}

We present now a sample disk galaxy formation simulation, undertaken
within a $\Lambda$CDM cosmological framework, in order to demonstrate
the capabilities of chemodynamical codes such as {\tt GCD+}.
We carried out a series of
high-resolution simulations within a 
$\Lambda$-dominated cold dark matter ($\Lambda$CDM) cosmology
($\Omega_0$=0.3, $\Lambda_0$=0.7, $\Omega_{\rm b}$=0.019$h^{-2}$, $h$=0.7,
$\sigma_8$=0.9). We used a multi-resolution technique (described
below) to achieve high-resolution in the region of interest, 
including the tidal forces from
large scale structures.  The multi-resolution initial conditions were
constructed using Bertschinger's (2001) publically available software 
{\tt GRAFIC2}. 
First, a low-resolution N-body \emph{only} simulation of a comoving
20$h^{-1}$ Mpc diameter sphere was performed; the 
mean separation of the particles therein was
20$h^{-1}$/64 Mpc. The mass of each particle was $3.63\times 10^{9} {\rm
M_{\odot}}$, and a fixed softening of 18.0~kpc was applied.
Next, at redshift $z=0$, we selected an $\sim$10 Mpc diameter spherical region
which contains several galaxy-sized DM halos.
We traced the particles which fall into the selected region
back to the initial conditions at $z=43.5$ and identified the volume 
which consists of those particles. Within this arbitrarily shaped volume,
we replaced the low resolution particles with particles a factor of 64 
times less massive. The initial density and velocities for the less massive 
particles are self-consistently calculated by GRAFIC2, taking into
account the density fields of the lower resolution region.
Finally, we re-simulated the full volume (20$h^{-1}$~Mpc
sphere), but now including all gas dynamics, cooling, and star
formation.\footnote{The gas component though was included only within the 
high-resolution region.}
The surrounding low-resolution region contributes to the high-resolution
region only through gravity. The mass and softening length of individual
gas (dark matter) particles in the high-resolution region were 
$7.33\times10^6$ ($4.94\times10^7$) ${\rm M}_{\rm \odot}$ and 1.14 (2.15) kpc,
respectively.

At $z=0$, using a friends-of-friends methodology, we identified six stellar
systems which consisted of more than 2000 star particles. Two of these
systems resembled large disk-like systems with kinematics consistent of 
rotational support; one was chosen as the target galaxy.  The total virial
mass of this galaxy is $\sim$2.4$\times$10$^{12}$~${\rm
M_{\odot}}$, and there exists a companion with $\sim$30\% the mass of
the target at a galactocentric distance of $\sim$270~kpc.
The virial mass is defined as the mass within the virial radius,
which itself is the radius of a sphere containing a mean density
of 178~$\Omega_0^{0.45}$ times the critical values ($\rho_{\rm crit}=3
{\rm H_0} /8\pi G$), after Eke, Navarro \& Frenk (1998).

Figure~5 shows the morphological evolution of the
dark matter in a central
portion of the simulation volume, and the evolution of the stellar
component in a 200~kpc region centred on the target galaxy.
The lower panels correspond to the predicted $J$-band (in the rest frame) 
image of the target galaxy.
In our simulations, the star particles each carry their own age and
metallicity ``tag'', due to the self-consistent chemodynamical nature of the
calculation.
This enables us to generate an optical-to-near
infrared spectral energy distribution
for the target galaxy (here, using the Kodama \& Arimoto 1997 
spectral synthesis package).

The spectral energy distribution of each star
particle is assumed to be that of a simple stellar population -
i.e. a coeval and chemically
homogeneous assembly of stars. We take into account the $k$-correction, but
do not consider here the effects of dust absorption. 
Figure~5 demonstrates
that the galaxy forms through conventional hierarchical clustering before
$z=1$; the disk has subsequently been built-up smoothly.

Figure~6 demonstrates the time evolution of the phase space
information of stars within a satellite which accretes
onto the central galaxy at $z\sim0.3$. 
At $z=0.34$, the tidal stream of the satellite appears.
The stream is identifiable in the radial velocity
($V_{\rm R,gal}$) -- galactocentric radius ($R_{\rm gal}$) diagram,
in addition to the $V_{\rm R,gal}$ -- rotational velocity
($V_{\rm \phi,gal}$) diagram.
This satellite is completely disrupted at $z=0$,
and the $V_{\rm R,gal}$--$R_{\rm gal}$ information of 
the member star particles of the satellite also disappear.
However, in the $V_{\rm R,gal}$--$V_{\rm \phi,gal}$ diagram,
the distribution of the member star particles
is similar to that at $z=0.34$. We confirmed
that the velocity phase space information of the member stars
is conserved very well, 
even after the satellite is spatially disrupted
(see also Helmi \& White 1999; Helmi \& de Zeeuw 2000).

It is worth noting however that it is difficult to identify
the member stars of the accreted satellites
in the velocity phase space diagram of the full sample of observed stars.
The red dots in Figure~6 represent the halo stars. 
Here, we define the star particles
with [Fe/H]$<-0.6$ (Chiba \& Beers 2000) as the halo stars. 
Figure~6 shows that in the velocity phase space diagram,
the distribution of the satellite member stars overlaps 
that of the halo stars.
Thus, to identify the member stars of the accreted satellites, 
additional observational information is required. 
Since the member stars formed within a small galaxy,
it is anticipated that they might each inherit a unique
chemical ``fingerprint'' (Freeman \& Bland-Hawthorn 2002).
Hence, the combination of such chemical tags and
kinematics can be a powerful tool to identify the field stars
which orginated within now accreted satellites. Comparing
such observations and the results of chemodynamical simulations will be
critical to understanding the formation history of the Milky Way.
Unfortunately, current numerical simulations still struggle to overcome
the classical ``over-cooling problem'' (White \& Frenk 1991), the signatures
of which are a high-redshift star formation rate in excess of that 
observed, and stellar halos which are both too massive and too 
metal-rich (Brook et~al. 2003a,b; Helmi et~al. 2003). 
The exact physical mechanism required to solve this problem
remains uncertain, although a framework predicated upon an enhanced
supernova feedback efficiency is one leading candidate
(Navarro \& Steinmetz 2000; Brook et~al. 2004).

\section{Spectrophotometry}

Determining the epoch of galaxy formation and understanding the consequent
chemical evolution are amongst the fundamental quests of modern cosmology. An 
complementary approach to addressing the chemical evolution of galaxies is via 
the use of spectrophotometry.
One can directly derive the age and metallicity of a
galaxy, and their respective gradients therein, by comparing its
observational integrated colours and/or its spectral line indices with
theoretical predictions from stellar population synthesis techniques.  The
integrated properties of star clusters around a galaxy can also be used
to induce their host galaxy's chemical evolutionary path.

Necessary ingredients for theoretical spectrophotometric predictions include
stellar evolutionary tracks, isochrones, and a corresponding
stellar atmosphere library which covers a wide range of stellar parameters,
such as metallicity, temperature, and surface gravity.
The stellar population synthesis models presented below, for example, are
based upon the $Y^{2}$ 
Isochrones\footnote{\tt http://csaweb.yonsei.ac.kr/$\sim$kim/yyiso.html\rm} 
(Kim et~al. 2002) with [$\alpha$/Fe]=$+$0.3, coupled to the 
post-red giant branch stellar evolutionary tracks of
Yi, Demarque \& Kim (1997). The stellar library of Lejeune, Cuisinier \&
Buser (1998) was taken for the conversion from theoretical quantities to
observable quantities.

Both age-sensitive and metallicity-sensitive spectrophotometric quantities
are initially constructed for a grid of simple stellar populations, for a
range of age and metallicity. Composite stellar population
spectrophotometric quantities can then be generated by convolving any
given star formation history (with the requisite self-consistent treatment
of chemical enrichment) with the grid of simple stellar population results
(weighted by the number of stars populating each stellar evolutionary
stage of each simple stellar population). The spectrophotometric
quantities calculated from these composite populations can then be compared
directly with observational data, and the age and metallicity of the
underlying stellar population extracted. Once we have a fair selection of
sample galaxies in terms of age and metallicity, we are able to
investigate the detailed abundance properties along the age sequence to
understand the chemical evolution of galaxies.

\begin{figure}
\begin{center}
  \leavevmode
  \epsfxsize 8cm
  \epsfysize 8cm
  \epsffile{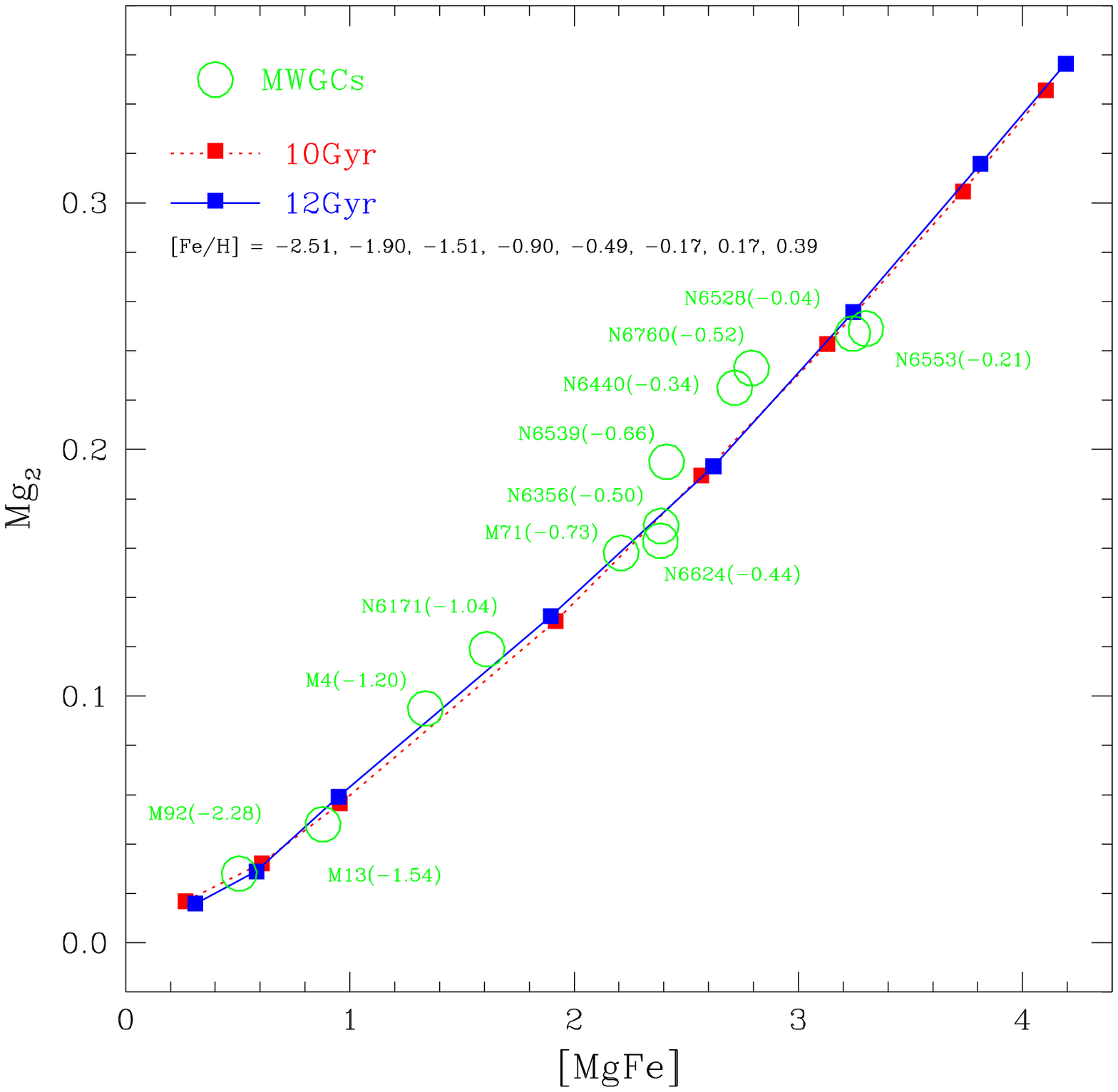}
\end{center}
\caption{The metallicity [Fe/H] of Milky Way globular clusters are
well-reproduced from this spectral index - index plot.  The 
data shown - including Mg$_2$ and [MgFe] indices - are
are taken from Cohen et~al. (1998) and the metallicities in parentheses are
from Harris (1996, Feb. 2003 Version).
}
\end{figure}

Figure~7 demonstrates the use of spectral index vs. index plots as tools
for estimating metallicity.  A sample of integrated spectra for
12 Milky Way globular clusters is shown (from Cohen et~al. 1998), which was
used to define a grid of Lick indices which are then compared with our
recent models (H.-c. Lee \& B.K. Gibson, in preparation). The [MgFe] index
is defined as $\sqrt{Mg b \times <Fe>}$, where $<Fe>$ is 
(Fe5270 + Fe5335)/2. It is
found that the metallicity [Fe/H] that is acquired independently from
the Harris compilation (Feb. 2003 Version) is surprisingly well recovered.  
This kind of calibration is necessary and should be quite useful for the
derivation of metallicity for extragalactic star clusters and/or external
galaxies that are not resolved into individual stars.

\begin{figure}
\begin{center}
  \leavevmode
  \epsfxsize 8cm
  \epsfysize 8cm
  \epsffile{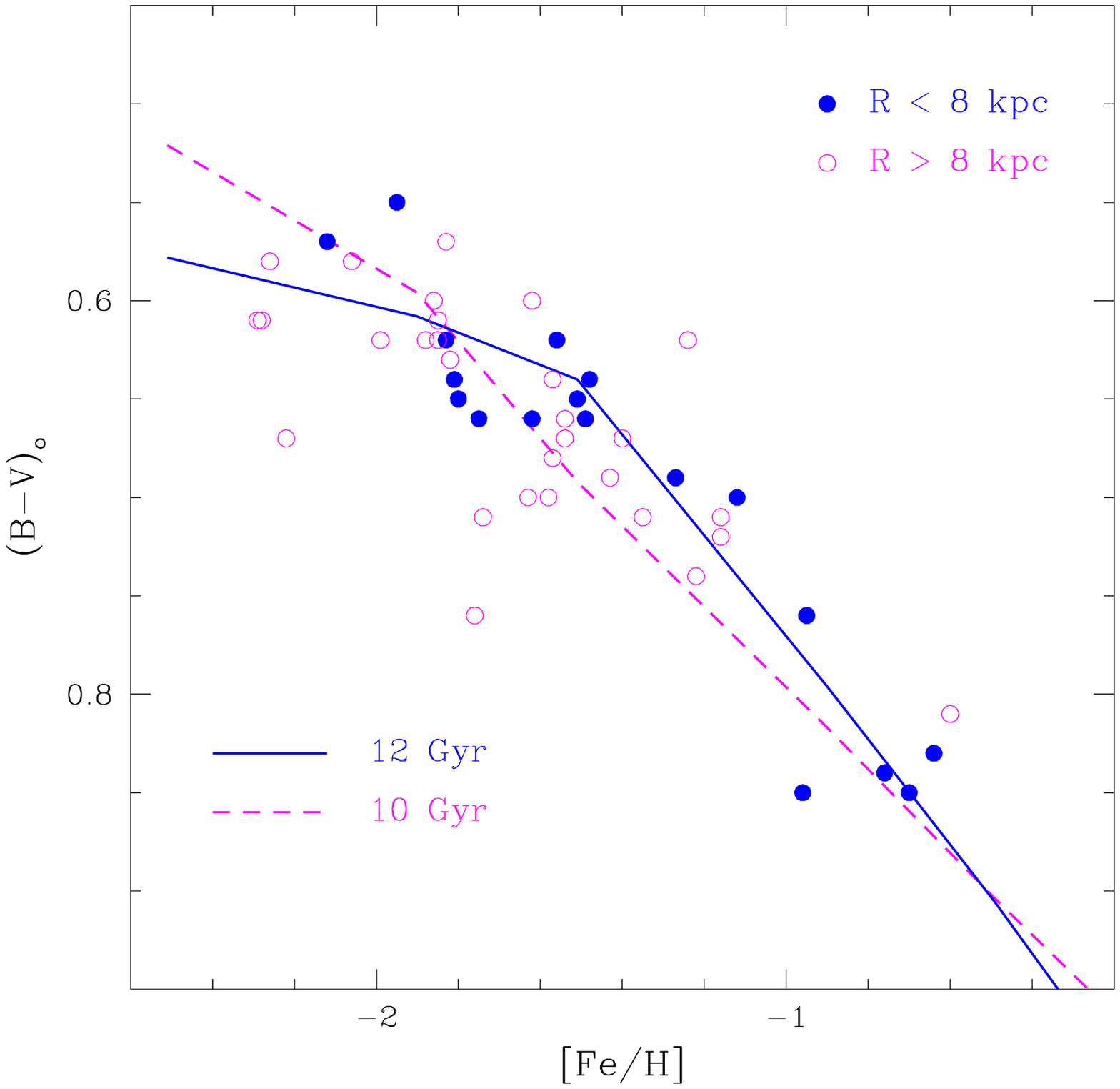}
\end{center}
\caption{The relatively low-reddened Milky Way globular clusters [$E$($B$
$-$ $V$) $<$ 0.2] are used to calibrate our models in the ($B$ $-$
$V$)$_{o}$ vs. [Fe/H] plane. The dashed and solid lines represent ages of
10 Gyr and 12 Gyr, respectively. Filled (open) circles correspond to inner
(outer) halo globulars.}
\end{figure}

Figure~8 shows the importance of the realistic manifestation of
horizontal-branch (HB) morphologies in the stellar population synthesis
models for relatively old stellar systems ($\tau$ $>$ 8 Gyr) (Lee,
Yoon \& Lee 2000; Lee, Lee \& Gibson 2002).  It appears that models
with a proper treatment of blue HB stars reproduce the differences
between inner and outer halo clusters,\footnote{The
inclusion of blue HB stars also 
led Lee et~al. (2000,2002) to suggest that giant
elliptical galaxies may be $\sim$1$-$3 billion years older than the Milky Way.}
in the sense that the inner halo
clusters are not only more tightly grouped along the isochrone than the
more scattered outer halo counterparts, but also relatively older. This is
interesting in that the outer halo clusters with the wider range of colours
at a given metallicity may indicate a different chemical ``origin'',
perhaps from different environments such as satellite dwarf
galaxies.

\begin{figure}
\begin{center}
  \leavevmode
  \epsfxsize 8cm
  \epsfysize 9cm
  \epsffile{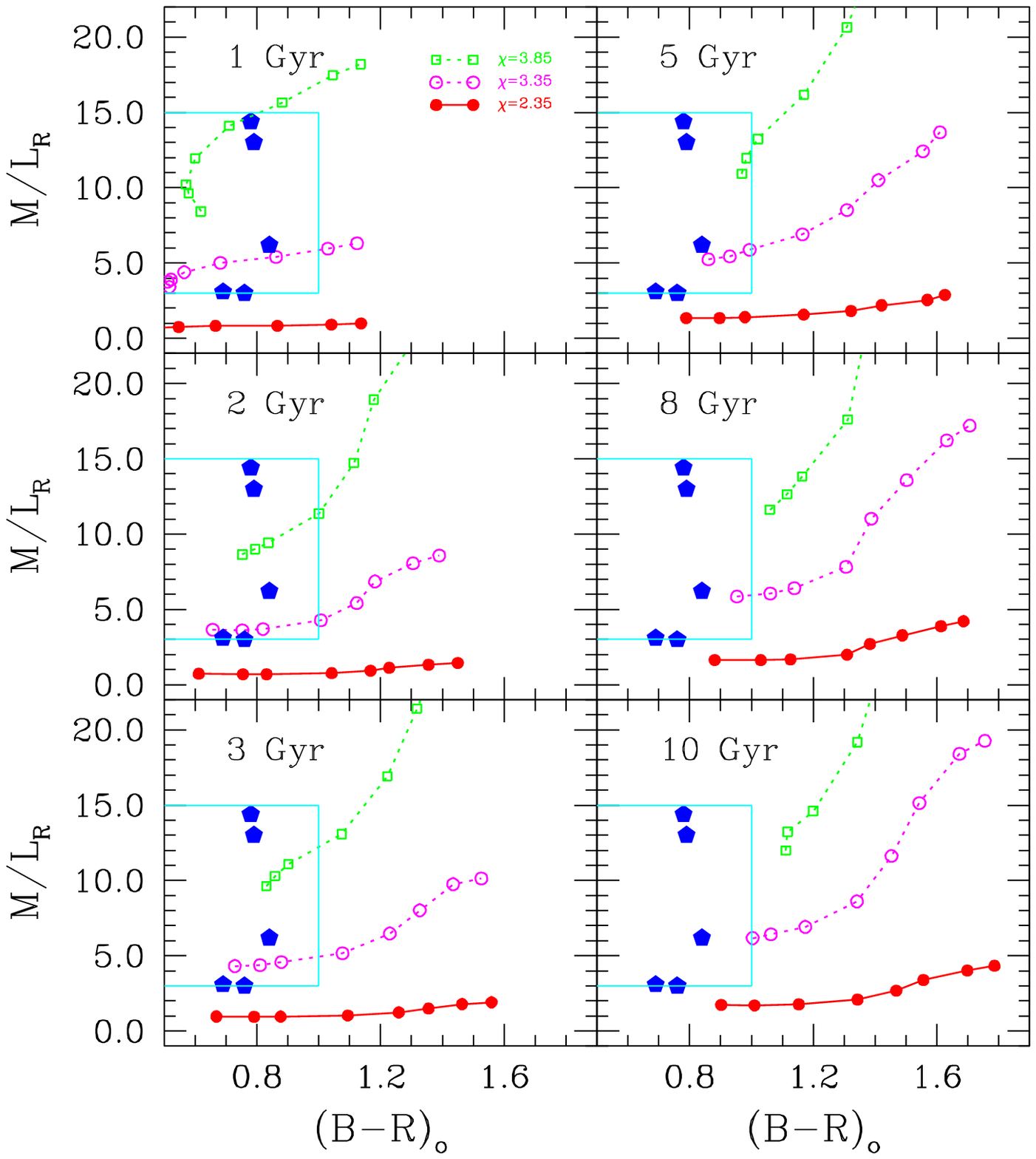}
\end{center}
\caption{Our simple stellar population predictions for $(B-R)_o$ vs.
$M/L_R$ as a function of age for three IMF slopes. The solid line
with filled circles corresponds to the standard Salpeter (1955)
IMF (by number) with an exponent
$\chi$=2.35. The dotted lines with open circles and open squares are
$\chi$=3.35 and $\chi$=3.85 cases, respectively. At a given age and 
IMF exponent, the metallicity 
[Fe/H] is $-$2.51, $-$1.90, $-$1.51, $-$0.90, $-$0.49,
$-$0.17, $+$0.17, and $+$0.39, from left to right. The filled pentagons are
Fuchs (2002) sample of LSB galaxies.
From the left panel, it is suggested that these LSB
galaxies with high $M/L_R$ ratios and
blue colours ($(B-R)_o$ $<$ 1.0) may be consistent with a relatively low
metallicity and rather recent ($\simlt$2~Gyr ago) star formation,
by introducing a steep IMF (Lee et~al. 2004).
}
\end{figure}

Another important parameter that controls spectrophotometric quantities, as
well as the mass-to-luminosity (M/L) ratio, is the 
IMF. A top-heavy IMF leads to highly efficient chemical enrichment
due to a preponderance of Type~II SNe.  Conversely, chemical enrichment
is minimised under the adoption of a bottom-heavy IMF.
A recent attempt by Fuchs (2002) to set constraints on the mass of the
disks of low surface brightness (LSB) galaxies by employing 
density wave theory is intriguing in this respect. Fuchs used a
sample of five LSBs with clear spiral structure to claim that
each possess surprisingly
high stellar mass-to-luminosity ratios in the R-band
($M/L_R$ $\geq$ 3), in addition to their
blue colours ($(B-R)_o$ $<$ 1.0). 
Figure~9 demonstrates that the range of colours and mass-to-luminosity
ratios spanned by the Fuchs LSB sample are consistent with a low-metallicity
([Fe/H]$\simlt$$-$1.5), recent ($\simlt$2\,Gyr), burst of star formation,
under the assumption of an IMF significantly steeper than that of
Salpeter (1955).

\section{Future Directions}

Any model for the formation and evolution of the Milky Way is only as good as
the observational data upon which it is calibrated.  
We wish to end this review with an outline of some of the exciting
observational programs which will come to fruition over the coming decade.
Each of these datasets is capable of constraining
- in a \emph{new and significant} manner - the GCE models discussed in 
\S~3 $-$ \S~6.

We are fortunate that for the Milky Way we can, in principle, obtain full
six-dimensional phase space (spatial and kinematical) and chemical
information, for individual stars.  Such data provides unique insights into the
detailed formation history of our Galaxy
(e.g. Eggen, Lynden-Bell \& Sandage 1962; Chiba \& Beers 2000;
Bekki \& Chiba 2001; Freeman \& Bland-Hawthorn 2002; Brook et~al. 2003a).
The discovery of the disrupting Sagittarius dwarf (Ibata et~al. 1994) and 
other halo substructure signatures (Helmi et~al. 1999; Chiba \& Beers 2000;
Gilmore, Wyse \& Norris 2002; Brook et~al. 2003a) has also demonstrated the
value of stellar kinematics in reconstructing satellite accretion events.
Such events can 
(potentially) aid in determining the fraction of the halo which was accreted
and the fraction which formed \it in situ \rm
(Helmi \& White 1999; Helmi \& de Zeeuw 2000; Harding et~al. 2001).

We identify three landmark projects which will advance significantly
the field of ``Galactic Archaeology'':

\begin{itemize}
\item \textbf{RAVE\footnote{\tt http://astronomy.swin.edu.au/RAVE/\rm} 
(RAdial Velocity Experiment)} is an ambitious
all-sky survey (complete to V=16) aimed at measuring the
radial velocities (with precision $\simlt$2\,km~s$^{-1}$), metallicities,
and abundance ratios (both to $\sim$0.1~dex precision) of 50
million stars using the United Kingdom Schmidt Telescope (UKST),
together with a northern counterpart, over the period
2006 - 2010, providing a \emph{vast} stellar kinematic and chemical
database.
A two-year pilot survey commenced on the UKST in April 2003, making
use of the existing 6dF Facility.  This pilot survey will obtain
comparable quality data to the main survey for 10$^5$ stars 
(spanning $\sim$8500\,deg$^2$, 9$<$I$<$12), of which approximately half 
have accurate Tycho-2 proper motions.
\item \textbf{ESA's GAIA\footnote{\tt http://astro.estec.esa.nl/GAIA/}
satellite mission}, scheduled
for launch in 2010, will conduct a census of $\sim$10$^9$
stars - $\sim$100 revisits per star over a five-year period -
measuring positions of all objects down to 
V=20.
Positional accuracies of $\sim$4\,$\mu$as at V=10 and $\sim$10\,$\mu$as 
at V=15 will yield distances accurate to 10\% at heliocentric distances
of $\sim$10\,kpc.  Radial velocity accuracy comparable to that of RAVE
is expected (also to V=16).
\item \textbf{NASA's Space Interferometry Mission - 
SIM\footnote{\tt http://sim.jpl.nasa.gov/\rm}},
scheduled for launch in 2009, complements GAIA's 4\,$\mu$as positional
accuracy with a narrow-angle mode allowing accuracies of 
$\sim$ 1\,$\mu$as.
\end{itemize}

There currently exists little compelling evidence that a massive 
($>$10$^{10}$\,M$_\odot$) satellite has been accreted by the Milky Way within
the past $\sim$10\,Gyr (Gilmore et~al. 2002).\footnote{The potential impact
that such a massive satellite would have upon the disk supports this
contention (Kawata et~al. 2003).}
Instead, later accreted satelites are more likely to resemble dwarf
systems such as the Local Group's dwarf spheroidals (dSphs).
Recent (and spectacular) high-resolution spectroscopic observations of
individual stars in dSphs have demonstrated that their ratio of 
$\alpha$-elements to iron tend to be close to solar 
(e.g. Shetrone et~al. 2001,2003). This elemental abundance 
pattern is \emph{not} reflected in the present-day halo field stars - a
somewhat surprising result considering the aforementioned kinematical
evidence for halo acccretion events.  An inescapable conclusion to be
drawn from these data is that the present-day Local Group dSphs are not
the primary ``stellar donors'' to the Galactic halo (Tolstoy et~al. 2003).
Whether such accreted satellites are responsible for the thick disk though
remains an intriguing possibility (Bekki \& Chiba 2002; Abadi et~al. 2003b).
Chemical ``tagging'' of Local Group dSphs is still very much in its infancy,
but a burgeoning field which will see enormous increases in sample sizes over
the coming few years, thanks to substantial investments of 8m-class time.

The hunt for the most primitive stars in the Milky Way has led to
some of the most exciting discoveries in recent astronomy, including
the detection of a star with an iron abundance less than 1/200,000 that
of the Sun (Christlieb et~al. 2002). Having formed from almost pristine
primordial gas, this star (HE0107-5240) is a nearby counterpart to the
high-redshift universe and provides insight into the earliest epochs
of Galaxy formation. 
%The term population III applies to the first
%generation of stars to have formed from zero-metallicity gas with
%``Big Bang'' composition. 
Two factors could prohibit the detection of
a bona fide
zero-metallicity Pop~III star: 1) metal-free gas might favour the
formation of higher mass stars with short lifetimes, such that there
are no surviving Pop~III stars today (Nakamura \& Umemura 2001); 
and 2) the surfaces of Pop~III
stars may have been polluted with metals either from internal
processing or through the accretion of interstellar matter 
(Shigeyama et~al. 2003). Despite
their paucity, thousands of candidate metal-poor stars have been
selected for follow-up spectroscopy by programs such as the HK and
Hamburg/ESO surveys (Beers 2000). 
The tally of $\sim$100 stars with [Fe/H]$<$$-$3
found by the HK survey is expected to grow by more than a factor of
five with the Hamburg/ESO program (Christlieb 2003). As alluded to
in \S~4, the abundance
pattern in extremely metal-poor stars may reflect the chemical
fingerprint of a single Pop~III star and provides empirical constraints on
models of ``The First Stars''
(e.g. Umeda \& Nomoto 2003). A project using
the Sloan Digital Sky Survey dataset to find metal-poor stars
in the halo and thick disk of the Milky Way is also underway 
(Allende~Prieto et~al. 2003).

We have only touched the ``tip of the iceberg'' here - the combination of
``big glass'' and ``big surveys'' over the next decade will see an 
explosion of chemo-kinematical data coming on-line.\footnote{Indeed,
the data ``glut'' that surveys like RAVE will contribute to is one 
motivation driving the development of virtual observatories.}  
Such a wealth of impending information makes this the optimal 
time to be a theorist interested in understanding the
detailed formation and evolution of our Milky Way Galaxy.

\section*{Acknowledgements}

We thank Chris Flynn, Stefan Harfst, Chris Brook, Chris Thom, Tim Connors, 
Alexander Knebe, Mike Beasley, Antonio Pipino,
Alessandro Chieffi, Marco Limongi, John Lattanzio, and
Amanda Karakas, for helpful discussions. We acknowledge the Yukawa
Institute Computer Facility, the Astronomical Data Analysis Centre of the
National Astronomical Observatory Japan, and the Victorian and Australian
Partnerships for Advanced Computing, for the use of their facilities
during the preparation of this review. This work has benefited from the 
ongoing financial support of the 
Australian Research Council, to whom we are indebted.

\section*{References}

\reference Abadi M.G., Navarro, J.F., Steinmetz, M. \& Eke, V.R. 2003a, ApJ, 591, 499
\reference Abadi M.G., Navarro, J.F., Steinmetz, M. \& Eke, V.R. 2003b, ApJ, 597, 21
\reference Alib\'es, A., Labay, J. \& Canal, R. 2001, A\&A, 370, 1103
\reference Allende Prieto, C., Lambert, D.L. \& Asplund, M. 2001, ApJ, 556, L63
\reference Allende Prieto, C., Beers, T.C., Li, Y., et~al. 2003,
Carnegie Observatories Astrophysics Series, Vol. 4: Origin and Evolution of 
the Elements, eds.  A. McWilliam and M. Rauch (Pasadena: Carnegie
Observatories,\hfill\break
\tt http://www.ociw.edu/ociw/symposia/series/\hfill\break
symposium4/proceedings.html\rm)
\reference Anders, E. \& Grevesse, N. 1989, GeCoA, 53, 197 
\reference Argast, D., Samland, M., Gerhard, O.E. \& Thielemann, F.-K. 2000, A\&A, 356, 873
\reference Argast, D., Samland, M., Thielemann, F.-K. \& Gerhard, O.E. 2002, A\&A, 388, 842
\reference Barnes, J.E. \& Hut P. 1986, Nature, 324, 446
\reference Beers, T.C. 2000, in Weiss, A., Abel, T.G. \& Hill, V., eds.,
The First Stars, Berlin:Springer, p.3
\reference Bekki, K. \& Chiba, M. 2001, ApJ, 558, 666
\reference Bekki, K. \& Chiba, M. 2002, ApJ, 566, 245
\reference Berczik, P. 1999, A\&A, 348, 371
\reference Bertschinger, E. 2001, ApJS, 137, 1
\reference Boissier, S. \& Prantzos, N. 2000, MNRAS, 312, 398
\reference Brook, C.B., Kawata, D., Gibson, B.K. \& Flynn C. 2003a, ApJL, 585, 125
\reference Brook, C.B., Kawata, D. \& Gibson, B.K. 2003b, MNRAS, 343, 913
\reference Brook, C.B., Kawata, D., Gibson, B.K. \& Flynn C. 2004, MNRAS, in
press
\reference Buser, R., Rong, J. \& Karaali, S. 1998, A\&A, 331, 934
\reference Carraro, G., Lia, C. \& Chiosi, C. 1998, MNRAS, 297, 1021
\reference Chang, R.X.,  Hou, J.L., Shu, C.G. \& Fu, C.Q. 1999, A\&A, 350, 38 
\reference Chen, B., Stoughton, C., Smith, J.A., et~al. 2001, ApJ, 553, 184
\reference Chiappini, C., Matteucci, F. \& Gratton, R. 1997, ApJ, 477, 765
\reference Chiappini, C., Matteucci, F. \& Padoan, P. 2000, ApJ, 528, 711
\reference Chiba, M. \& Beers, T.C. 2000, AJ, 119, 2843
\reference Chieffi, A. \& Limongi, M. 2002, ApJ, 577, 281
\reference Christlieb, N. 2003, Rev. Mod. Astron., 16, 191
\reference Christlieb, N., Bessell, M.S., Beers, T.C., et~al., 2002, Nature, 419, 904
\reference Cohen, J.G., Blakeslee, J.P. \& Ryzhov, A. 1998, ApJ, 496, 808 
\reference Copi, C.J. 1997, ApJ, 487, 704 
\reference Dekel, A. \& Silk, J. 1986, ApJ, 303, 39 
\reference Dopita, M.A. \&  Ryder, S.D. 1994, ApJ, 430, 163 
\reference Edvardsson, B., Andersen, J., Gustafsson, B., Lambert, D.L., Nissen, P.E. \& Tomkin, J.\ 1993, A\&A, 275, 101
\reference Eggen, O.J., Lynden-Bell, D. \& Sandage, A.R. 1962, ApJ, 136, 748
\reference Eke, V.R., Navarro, J.F. \& Frenk, C.S. 1998, ApJ, 503, 569
\reference Feltzing, S., Holmberg, J. \& Hurley, J.R. 2001, A\&A, 377, 911 
\reference Fenner, Y. \& Gibson,  B.K. 2003, PASA, 20, 189 
\reference Fenner, Y., Gibson, B.K., Lee, H.-c., Karakas, A.I., Lattanzio, J.C., Chieffi, A., Limongi, M. \& Yong, D. 2003, PASA, 20, 340
\reference Freeman, K.C. \& Bland-Hawthorn, J. 2002, ARAA, 40, 487
\reference Fuchs B., 2002, in Klapdor-Kleingrothaus H.V. \& Viollier R.D., 
 eds., Dark Matter in Astro- and Particle Physics, Berlin:Springer, p.28 
\reference Gay, P.L. \& Lambert, D.L. 2000, ApJ, 533, 260 
\reference Gibson, B.K. 1997, MNRAS, 290, 471
\reference Gibson, B.K. \& Mould, J.R. 1997, ApJ, 482, 98
\reference Gibson, B.K., Loewenstein, M. \& Mushotzky, R.F. 1997, MNRAS, 290, 623
\reference Gilmore, G., Reid, N. 1983, MNRAS, 202, 1025
\reference Gilmore, G., Wyse, R.F.G. \& Kuijken, K. 1989, ARA\&A, 27, 555
\reference Gilmore, G., Wyse, R.F.G. \& Norris, J.E. 2002, ApJ, 574, L39
\reference Gingold, R.A. \& Monaghan, J.J. 1977, MNRAS, 181, 375
\reference Greggio, L. \& Renzini, A. 1983, A\&A, 118, 217
\reference Harding, P., Morrison, H.L., Olszewski, E.W., Arabadjis, J., Mateo, M., et al. 2001, AJ, 122, 1397
\reference Harris, W.E. 1996, AJ, 112, 1487
\reference Helmi, A. \& de Zeeuw, P.T. 2000, MNRAS, 319, 657
\reference Helmi, A. \& White, S.D.M. 1999, MNRAS, 307, 495
\reference Helmi, A., White, S.D.M., de Zeeuw, P.T. \& Zhao, H. 1999, Nature, 402, 53
\reference Helmi, A., Navarro, J.F., Meza, A., Steinmetz, M. \& Eke, V.R. 2003,
ApJ, 592, L25
\reference Holweger, H. 2001, in Wimmer-Schweingruber, R.F., eds., Solar and Galactic Composition, AIP, p.23
\reference Ibata, R., Gilmore, G. \& Irwin, M.J. 1994, Nature, 370, 194
\reference Ibukiyama, A. \& Arimoto, N. 2002, A\&A, 394, 927
\reference Ikuta, C. \& Arimoto, N. 1999, PASJ, 51, 459 
\reference Iwamoto, K., Brachwitz, F., Nomoto, K., Kishimoto, N., Umeda, H., Hix, W.R. \& Thielemann, F.-K. 1999, ApJS, 125, 439
\reference Karakas, A.I. \& Lattanzio, J.C., 2003, PASA, 20, 279
\reference Kawata, D. 1999, PASJ, 51, 931
\reference Kawata, D. \& Gibson B.K. 2003a, MNRAS, 340, 908
\reference Kawata, D. \& Gibson B.K. 2003b, MNRAS, 346, 135
\reference Kawata, D., Thom, C. \& Gibson, B.K. 2003, PASA, 20, 263
\reference Katz, N., Weinberg, D.H. \& Hernquist, L. 1996, ApJS, 105, 19
\reference Kay, S.T., Pearce, F.R., Jenkins, A., Frenk, C.S., White, S.D.M., Thomas, P.A. \& Couchman, H.M.P. 2000, MNRAS, 316, 374
\reference Kim Y.-C., Demarque P., Yi S.K. \& Alexander D.R. 2002, ApJS, 143, 499
\reference Kobayashi, C., Tsujimoto, T. \& Nomoto, K. 2000, ApJ, 539, 26
\reference Koda, J., Sofue, Y. \& Wada K. 2000, ApJL, 531, 17 
\reference Kodama, T. \& Arimoto, N. 1997, A\&A, 320, 41
\reference Kotoneva, E., Flynn, C., Chiappini, C. \& Matteucci, F. 2002, MNRAS, 336, 879
\reference Kroupa, P., Tout, C.A. \& Gilmore, G. 1993, MNRAS, 262, 545 
\reference Larson, R.B. 1974, MNRAS, 169, 229 
\reference Larson, R.B. 1976, MNRAS, 176, 31 
\reference Lee, H.-c., Lee, Y.-W. \& Gibson, B.K. 2002, AJ, 124, 2664
\reference Lee, H.-c., Yoon, S.-J. \& Lee, Y.-W. 2000, AJ, 120, 998
\reference Lee, H.-c., Gibson, B.K., Flynn, C., Kawata, D. \& Beasley, M.A.
2004, MNRAS, submitted
\reference Lejeune T., Cuisinier F. \& Buser R. 1998, A\&AS, 130, 65
\reference Lucy, L.B. 1977, AJ, 82, 1013
\reference Maeder, A. 1992, A\&A, 264, 105
\reference Malinie, G., Hartmann, D.H. \& Mathews, G.J. 1991, ApJ, 376, 520 
\reference Malinie, G., Hartmann, D.H., Clayton, D.D. \& Mathews, G.J. 1993, ApJ, 413, 633 
\reference Matteucci, F. \& Greggio, L. 1986, A\&A, 154, 279 
\reference Matteucci, F. 2001, The Chemical Evolution of the Galaxy, 
Dordrecht:Kluwer
\reference Matteucci, F., Raiteri, C.M., Busson, M., Gallino, R. \& Gratton, R. 1993, A\&A, 272, 421 
\reference Matteucci, F. \& Tornamb\'e, A.\ 1987, A\&A, 185, 51 
\reference Mishenina, T.V., Kovtyukh, V.V., Soubiran, C., Travaglio, C. \& 
Busso, M. 2002, A\&A,  396, 189
\reference Nakamura, F. \& Umemura, M. 2001, ApJ, 548, 19
\reference Nakasato, N. \& Nomoto, K. 2003, ApJ, 588, 842
\reference Navarro, J.F. \& Steinmetz, M. 2000, ApJ, 538, 477
\reference Oey, M.S. 2000, ApJ, 542, L25
\reference Oey, M.S. 2003, MNRAS, 339, 849
\reference Oey, M.S. \& Clarke, C.J. 1997, MNRAS, 289, 570
\reference Pagel, B.E.J. \& Patchett, B.E. 1975, MNRAS, 172, 13
\reference Pilyugin, L.S. \& Edmunds, M.G. 1996, A\&A, 313, 792 
\reference Portinari, L. \& Chiosi, C. 1999, A\&A, 350, 827
\reference Portinari, L. \& Chiosi, C. 2000, A\&A, 355, 929
\reference Raiteri, C.M., Villata, M. \& Navarro, J.F. 1996, A\&A, 315, 105
\reference Recchi, S., Matteucci, F. \& D'Ercole, A. 2001, MNRAS, 322, 800
\reference Robin, A.C., Reyl\'e, C., Derri\`ere, S. \& Picaud, S. 2003, A\&A,
409, 523
\reference Rocha-Pinto, H.J., Scalo, J., Maciel, W.J. \& Flynn, C. 2000a, ApJL, 531, L115 
\reference Rocha-Pinto, H.J., Maciel, W.J., Scalo, J. \& Flynn, C. 2000b, A\&A, 358, 850 
\reference Salpeter E.E. 1955, ApJ, 121, 161
\reference Samland, M., Hensler, G. \& Theis, Ch. 1997, ApJ, 476, 544
\reference Scalo, J.M. 1986, Fund. Cosm. Phys., 11, 1
\reference Schmidt, M. 1959, ApJ, 129, 243
\reference Searle, L. \& Zinn, R. 1978, ApJ, 225, 357 
\reference Shetrone, M.D., C\^ot\'e, P. \& Sargent, W.L.W. 2001, ApJ, 548, 592
\reference Shetrone, M.D., Venn, K.A., Tolstoy, E., Primas, F., Hill, V. \& Kaufer, A. 2003, AJ, 125, 684
\reference Shigeyama, T. \& Tsujimoto, T. 1998, ApJL, 507, L135 
\reference Shigeyama, T., Tsujimoto, T. \& Yoshii, Y. 2003, ApJ, 586, L57
\reference Smartt, S.J. \& Rolleston, W.R.J. 1997, 481, L47
\reference Smecker-Hane, T.A. \& Wyse, R.F.G. 1992, AJ, 103, 1621
\reference Spite, M., DePagne, E., Cayrel, R., et~al. 2003,
Carnegie Observatories Astrophysics Series, Vol. 4: Origin and Evolution of 
the Elements, eds.  A. McWilliam and M. Rauch (Pasadena: Carnegie
Observatories,\hfill\break
\tt http://www.ociw.edu/ociw/symposia/series/\hfill\break
symposium4/proceedings.html\rm)
\reference Steinmetz, M. \& Muller, E. 1995, MNRAS, 276, 549
\reference Steinmetz, M \& Navarro, J.F. 1999, ApJ, 513, 555
\reference Sutherland, R.S. \& Dopita, M.A. 1993, ApJS, 88, 253
\reference Suzuki, T.K. \& Yoshii, Y. 2001, ApJ, 549, 303
\reference Timmes, F.X., Woosley, S.E. \& Weaver, T.A. 1995, ApJS, 98, 617
\reference Tinsley, B.M. 1980, Fund. Cosm. Phys., 5, 287
\reference Tolstoy, E., Venn, K.A., Shetrone, M.D., Primas, F., Hill, V., Kaufer, A. \& Thomas, S. 2003, AJ, 125, 707
\reference Travaglio, C., Galli, D. \& Burkert, A. 2001, ApJ, 547, 217 
\reference Travaglio, C., Kifonidis, K. \& M\"uller, E. 2003,
Carnegie Observatories Astrophysics Series, Vol. 4: Origin and Evolution of 
the Elements, eds.  A. McWilliam and M. Rauch (Pasadena: Carnegie
Observatories,\hfill\break
\tt http://www.ociw.edu/ociw/symposia/series/\hfill\break
symposium4/proceedings.html\rm)
\reference Tsujimoto, T. \& Shigeyama, T. 2002, ApJ, 571, L93
\reference Tsujimoto, T., Shigeyama, T. \& Yoshii, Y. 1999, ApJ, 519, L63
\reference Tsujimoto, T., Shigeyama, T. \& Yoshii, Y. 2002, ApJ, 565, 1011
\reference Twarog B.A., 1980, ApJ, 242, 242
\reference Umeda, H. \& Nomoto, K. 2002, ApJ, 565, 385
\reference Umeda, H. \& Nomoto, K. 2003, Nature, 422, 871
\reference van den Hoek, L.B. \& de Jong, T. 1997, A\&A, 318, 231 
\reference van den Hoek, L.B. \& Groenewegen, M.A.T. 1997, A\&AS, 123, 305
\reference van der Kruit, P.C. \& Searle, L. 1982, A\&A, 110, 79 
\reference White, S.D.M. \& Frenk, C.S. 1991, ApJ, 379, 52
\reference Woosley, S.E. \& Weaver, T.A. 1995, ApJS, 101, 181
\reference Worthey, G. \& Espana, A.L. 2003,
Carnegie Observatories Astrophysics Series, Vol. 4: Origin and Evolution of 
the Elements, eds.  A. McWilliam and M. Rauch (Pasadena: Carnegie
Observatories,\hfill\break
\tt http://www.ociw.edu/ociw/symposia/series/\hfill\break
symposium4/proceedings.html\rm)
\reference Wyse, R.F.G. \& Gilmore, G. 1988, AJ, 95, 1404
\reference Wyse, R.F.G. \& Gilmore, G. 1993, ASP Conf.~Ser.~ 48: The Globular
Cluster-Galaxy Connection, 727
\reference Yi S., Demarque P. \& Kim Y.-C. 1997, ApJ, 482, 677
\reference Yoshii, Y. \& Rodgers, A.W. 1989, AJ, 98 853

\end{document}